\documentclass{aa}  

\usepackage{graphicx}
\usepackage{txfonts}
\usepackage{hyperref}
\usepackage{xcolor}
\usepackage{orcidlink}
\usepackage{placeins}

\begin{document}

   \title{The compact jets of the black hole X-ray binary Swift~J1753.5--0127 during its 2023-2024 outburst}

   \subtitle{}

   \author{N.~Grollimund\inst{1}\fnmsep\thanks{Corresponding author: \texttt{\href{mailto:noa.grollimund@u-pariscite.fr}{noa.grollimund@u-pariscite.fr}}}\orcidlink{0009-0005-4358-5146}
          \and
          S.~Corbel\inst{1}\orcidlink{0000-0001-5538-5831}
          \and
          L.~Rhodes\inst{2,3}\orcidlink{0000-0003-2705-4941}
          \and
          D.~M.~Russell\inst{4}\orcidlink{0000-0002-3500-631X}
          \and
          M.~C.~Baglio\inst{5}\orcidlink{0000-0003-1285-4057}
          \and
          F.~Carotenuto\inst{6}\orcidlink{0000-0002-0426-3276}
          \and
          T.~Bouchet\inst{7}\orcidlink{0009-0003-5878-0978}
          \and
          P.~Saikia\inst{8}\orcidlink{0000-0002-5319-6620}
          \and
          R.~P.~Fender\inst{9,10}\orcidlink{0000-0002-5654-2744}
          \and
          S.~E.~Motta\inst{5,9}\orcidlink{0000-0002-6154-5843}
          \and
          P.~Casella\inst{6}\orcidlink{0000-0002-0752-3301}
          \and
          K.~Alabarta\inst{5}\orcidlink{0000-0003-0168-9906}
          \and
          F.~Lewis\inst{11,12}\orcidlink{0000-0003-3352-2334}
          \and
          A.~K.~Hughes\inst{9}\orcidlink{0000-0003-0764-0687}
          \and
          F.~J.~Cowie\inst{9}\orcidlink{0009-0009-0079-2419}
          \and
          J.~Crook-Mansour\inst{9}\orcidlink{0000-0001-7466-1192}
          \and
          X.~Zhang\inst{13}\orcidlink{0000-0002-8086-4049}
          }

   \institute{
   Université Paris Cité, Université Paris-Saclay, CEA, CNRS, AIM, F-91191 Gif-sur-Yvette, France
   \and
   Department of Physics, McGill University, 3600 rue University, Montréal, QC H3A 2T8, Canada
   \and
   Trottier Space Institute, McGill University, 3550 rue University, Montréal, QC H3A 2A7, Canada
   \and
   Center for Astrophysics and Space Science, New York University, Abu Dhabi, PO Box 129188, Abu Dhabi, UAE
   \and
   INAF--Osservatorio Astronomico di Brera, Via Bianchi 46, I-23807 Merate (LC), Italy
   \and
   INAF--Osservatorio Astronomico di Roma, Via Frascati 33, I-00078, Monte Porzio Catone (RM), Italy
   \and
   Julius-Maximilians-Universität Würzburg, Fakultät für Physik und Astronomie, Institut für Theoretische Physik und Astrophysik, Lehrstuhl für Astronomie, Emil-Fischer-Str. 31, 97074 Würzburg, Germany
   \and
   Department of Astronomy, Yale University, PO Box 208101, New Haven, CT 06520-8101, USA
   \and
   Astrophysics, Department of Physics, University of Oxford, Keble Road, Oxford, OX1 3RH, UK
   \and
   Department of Astronomy, University of Cape Town, Private Bag X3, Rondebosch 7701, South Africa
   \and
   Faulkes Telescope Project, Cardiff, Wales, UK
   \and
   The Schools' Observatory, Astrophysics Research Institute, Liverpool John Moores University, 146 Brownlow Hill, Liverpool L3 5RF, UK
   \and
   Shanghai Astronomical Observatory, Chinese Academy of Sciences, Shanghai 200030, China
   \\
             }

   \date{Received XXX; accepted XXX}

  \abstract 
   {X-ray binaries are excellent laboratories to study the coupling between accretion and ejection phenomena around compact objects through multiwavelength observations. Swift~J1753.5--0127 is a black hole low-mass X-ray binary discovered in outburst in 2005 that remained active for 11 years, followed by two mini-outbursts. This X-ray transient is known for exhibiting hard-only outbursts and was one of the first outliers of the radio/X-ray correlation.}
   {We investigate the behavior of the compact jets of Swift~J1753.5--0127 during its 2023-2024 outburst.}
   {We report a dense multiwavelength monitoring in the radio (MeerKAT, AMI--LA), NIR/optical/UV (REM, LCO, ATLAS, \textit{Swift}/UVOT) and X-ray (\textit{Swift}/XRT, NICER/XTI, INTEGRAL/IBIS) bands. Using new and archival luminosity measurements, we construct a refined, smoother radio/X-ray plane that covers the steep/outlier and shallow/transition branches of the radio/X-ray correlation. We also present spectral energy distributions (SEDs) constructed using quasi-simultaneous observations from four epochs.}
   {Swift~J1753.5--0127 underwent a faint, hard-only, double-peaked outburst, followed by a re-brightening in the optical. The updated radio/X-ray plane offers a clear view on the region occupied by this source, similar to MAXI J1348--630, but at lower radio luminosities than H~1743--322, and we confirm its hybrid behavior. The broad-band SED fitting reveals that a cool, truncated accretion disk dominates the NIR/optical/UV band. We find that the compact jets have a variable but weak contribution in the NIR, with a break frequency $\nu_\mathrm{break} \lesssim 5 \times 10^{12}~\mathrm{Hz}$ ($\lesssim 2 \times 10^{13}~\mathrm{Hz}$ if they do not contribute to the hard X-rays), in line with previous studies, providing additional evidence that radio-quiet sources are also infrared-quiet. We also find evidence that the radio spectral index evolves with X-ray luminosity, transitioning from inverted values near quiescence to flatter or negative values at higher luminosities. In the context of adiabatic jet models, this trend may indicate that the jet geometry becomes increasingly confined as the accretion rate rises.}
   {}

   \maketitle

\section{Introduction}
Black hole (BH) X-ray binaries (XRBs) are binary systems that consist of a stellar-mass BH orbited by a stellar companion. In low-mass systems, the star transfers material to the compact object by Roche lobe overflow via a hot accretion disk \citep[see][]{BahramianDegenaar2023}. While X-ray binaries typically spend most of their time in quiescence in which they accrete matter at very low X-ray luminosity, they are generally discovered in outburst, when the X-ray flux increases by several orders of magnitude. XRBs radiate all across the electromagnetic spectrum, ranging from the radio waves, emitted from the jets \citep[e.g.,][]{Fender2002}, to the X-rays, produced by the accretion flow \citep{Shakura_Sunyaev1973}, and possibly the base of the jets \citep[e.g.,][]{MarkoffNowakWilms2005}. During an outburst, transitions between several accretion states, defined by distinct X-ray spectral and timing signatures \citep[e.g.,][]{HomanBelloni2005, Remillard_McClintock2006, Belloni_Motta2016}, can occur as the system follows an hysteresis cycle. At the beginning of the outburst, the system is usually observed in the so-called hard state, whose X-ray spectrum is dominated by a non-thermal power-law component thought to arise from the inverse-Compton scattering of soft X-ray photons from the disk by a corona of hot electrons \citep[e.g.,][]{ThornePrice1975, Shapiro1976, Zdziarski_Gierlinski2004}. In this state, steady, compact jets, emitting partially self-absorbed synchrotron radiation, are typically observed at radio wavelengths \citep{Corbel2000, Fender2001}, resulting in a flat or slightly inverted spectrum ($\alpha \ga 0$, where the radio flux density follows $S_\nu \propto \nu^\alpha$). After spending days to months in the hard state, the system may transition to the soft state, which is characterized by a thermal soft X-ray emission from a geometrically thin, optically thick accretion disk. During such a transition, transient jets, that are bipolar, relativistic plasma bubbles, can be launched \citep[e.g.,][]{Fender1999, Miller-Jones2012}. These discrete ejecta, which have been associated with radio flares \citep[e.g.,][]{Hjellming1995, Bright2020}, display optically-thin synchrotron spectra \citep[$\alpha \sim -0.6$; e.g.,][]{Tetarenko2017}. In the soft state, the X-ray spectrum is dominated by the thermal emission of the disk and the compact jets are strongly quenched by more than 3 orders of magnitude \citep[e.g.,][]{Fender1999b, Coriat2011, Russell2019, Maccarone2020}. Some weeks to months later, the luminosity decreases, before the system transitions back to the hard state while the compact jets gradually re-activate \citep{Corbel2013b}, and finally goes back to quiescence.

The fundamental coupling between accretion and ejection in BH XRBs is manifested by various phenomenological behaviors. One of them is the correlation between the radio and X-ray luminosities ($L_\mathrm{R}$, $L_\mathrm{X}$) in the hard state. This non-linear relation, of the form of $L_\mathrm{R} \propto L_\mathrm{X}^\beta$, was originally discovered in GX~339--4 \citep{Hannikainen1998, Corbel2000, Corbel2003} and shortly after extended to other BH systems \citep{Gallo2003}. The $\beta$ index was found in the range $\sim 0.5$--$0.7$ depending on the source \citep{Corbel2003, Gallo2003, Gallo2006} and this correlation was labeled as the standard track. The correlation extends down to the quiescent state \citep[e.g.,][]{Gallo2006, Corbel2008, Corbel2013, Carotenuto2022b}. Finally, \citet{RussellD2006} showed that, the optical/infrared luminosity of BH systems is correlated with their X-ray luminosity, following a power-law relation with an index consistent with the established radio/X-ray correlation. Since the discovery of the standard track, a growing number of sources have been found to display a more \textit{radio-quiet} behavior and lie below this track \citep[e.g.,][]{Corbel2004, Coriat2011}, with a steeper correlation index ($\beta \gtrsim 1$). Some of these outlier sources have been observed to rejoin the standard track at lower luminosities, along a shallow transition branch, and have been labeled as hybrid \citep[e.g.,][]{Carotenuto2021b}. Recently, \citet{Crook-Mansour2026b} compiled the largest homogeneous radio/X-ray plane for black hole and neutron star XRBs to date, using MeerKAT and \textit{Swift}/XRT observations.

Swift~J1753.5--0127 (hereafter Swift~J1753) is a low-mass X-ray binary discovered in outburst by \textit{Swift}/BAT in 2005 \citep{Palmer2005} at relatively high Galactic latitude ($+12\degr$). Detections of counterparts in UV \citep{Still2005ATel}, optical \citep{Halpern2005ATel}, NIR \citep{Torres2005ATel} and radio \citep{Fender2005ATel} followed the same year. Based on the detection of quasi-periodic oscillations \citep{Morgan2005ATel} and the presence of a hard X-ray tail \citep{CadolleBel2007}, the accretor was classified as a BH candidate. This was strengthened later on by the determination of the orbital period of the system \citep[$P_\mathrm{orb} = 3.2443 \pm 0.0010~\mathrm{hrs}$; ][]{Zurita2008} and a lower limit on the mass of the compact object \citep[$M > 7.4 \pm 1.2~M_\sun$; ][]{Shaw2016b}. 

After 11 years in outburst, the source started to decay around September 2016 \citep{Russell2016ATel} and finally went back to quiescence two months later \citep{Shaw2016ATel}. During its long-term period of activity, Swift~J1753 mostly remained in the hard accretion state, while occasionally transitioning to the hard intermediate state before returning to the hard state \citep{Soleri2013}. An exception to these failed transitions occurred in 2015 when the source transitioned to a very low-luminosity soft state \citep{Shaw2015ATel, Shaw2016a}. In February and April 2017, the system underwent two consecutive mini-outbursts \citep{AlQasim2017ATel, Bernardini2017ATel} and returned to quiescence by July 2017 \citep{Zhang2017ATel}. The long duration and hard-only nature of the discovery outburst enabled extensive study of the behavior of the source in the radio/X-ray luminosity plane \citep{CadolleBel2007, Soleri2010, Rushton2016, Plotkin2017}. In fact, Swift~J1753 was one of the first outliers, that did not follow the standard track of the radio/X-ray correlation.

The distance to Swift~J1753 is poorly constrained. \citet{CadolleBel2007} suggested a probable distance between 4 and 8 kpc using an argument based on the comparison between the  measured $N_\mathrm{H}$ and the Galactic value in the line of sight. Another range of 2.5--7.9 kpc was found by \citet{Zurita2008}, from the peak X-ray luminosity during the initial outburst of the source. The measurement of optical and radio parallaxes (with low signal-to-noise ratios) led to astrometric distances of $7.15_{-2.48}^{+3.99}~\mathrm{kpc}$ and $8.8_{-4.0}^{+12}~\mathrm{kpc}$ \citep{Gandhi2019, Atri2019}. Recently, \citet{Yanes-Rizo2025} used optical observations of Swift~J1753 in quiescence to determine an inclination of $i = 79 \pm 5~\mathrm{deg}$, a black hole mass of $M = 8.8 \pm 1.3 M_\sun$ and a significantly lower refined distance of $D = 3.9 \pm 0.7~\mathrm{kpc}$.

On 2023 September 28, a new optical outburst of Swift~J1753 was detected using the Las Cumbres Observatory (LCO) network \citep{Alabarta2023ATel}, as part of the XB-NEWS monitoring program \citep{Russell2019_XBNEWS}, when the optical magnitude in the $i'$ band significantly increased compared to the quiescence level. The outburst was confirmed by X-ray and radio detections soon after \citep{Homan2023ATel, Grollimund2023ATel}. We initiated weekly monitoring of the source with the MeerKAT telescope as part of the X-KAT large open-time program, from 2023 October 1 until the end of the outburst by early March 2024 \citep{Alabarta2024ATel}. Within a few days after the return to quiescence, the system entered a second phase of activity that triggered the XB-NEWS alert \citep{Saikia2024ATel}, allowing us to schedule early follow-up radio observations \citep{Grollimund2024ATel16568}. Swift J1753 finally returned to quiescence in September 2024, as seen by LCO \citep{Alabarta2024ATel16818}.

In this work, we present multiwavelength observations of Swift~J1753 during its 2023-2024 outburst. This paper is structured as follows. Section~\ref{sect:obs} introduces the observations and data reduction we carried out, Sect.~\ref{sect:results} details the evolution of the outburst, an updated radio/X-ray plane for this source and the modeling of the broad-band SEDs. Section~\ref{sect:discussion} offers a discussion of the results and Sect.~\ref{sect:conclusions} presents our conclusions.

\section{Observations and data reduction}
\label{sect:obs}

\subsection{Radio}

\subsubsection{MeerKAT}
\label{sec:obs_meerkat}

Swift~J1753 was monitored at radio frequencies with the MeerKAT telescope \citep{JonasMeerKATteam2016} on a weekly cadence as part of the X-KAT large MeerKAT open-time program (Proposal ID: SCI-20230907-RF-01). Observations were carried out at a central frequency of 1.284 GHz (\textit{L}-band) with a total bandwidth of 856 MHz. We used PKS J1939--6342 as a primary calibrator and J1733--1304 as a phase calibrator. Each run consisted of a single $15$-min scan or several $30$-min scans of Swift~J1753 interleaved by 2-min scans of the phase calibrator. The primary calibrator was observed for 5 min at the start and the end of each run. When the source became very faint at the end of the second outburst, we carried out deeper observations (up to 2 hours, see Table \ref{tab:obs_meerkat}) to follow the source down to the lowest levels of emission in outburst. We reduced and imaged the data using the \textsc{oxkat\footnote{\url{https://github.com/IanHeywood/oxkat}}} reduction scripts \citep{oxkat}. The visibilities were initially flagged and calibrated using the Common Astronomy Software Application package \citep[\textsc{casa};][]{casa}. Additional flagging on the target data was conducted using \textsc{tricolour} \citep{tricolour}. For imaging the field of Swift~J1753, we used \textsc{wsclean} \citep{wsclean} with a Briggs weighting scheme (typical resolution of $\sim 7''$). After a step of direction-independent self-calibration with the \textsc{cubical} package \citep{cubical}, the target was imaged a second time using masked deconvolution. To extract the positions and flux densities, we fitted the source in the image plane with an elliptical Gaussian model.

\subsubsection{AMI--LA}

The Arcminute Microkelvin Imager - Large Array (AMI--LA) is an eight-dish interferometer, located near Cambridge (UK), with a resolution of $\sim 30''$. Observations with AMI--LA are made at 15.5 GHz with a bandwidth of 5 GHz \citep{Zwart2008, Hickish2018}. We observed the source between October 2023 and August 2024, with a cadence of $\sim$7\,days. The data were reduced using a custom pipeline \textsc{reduce\_dc} which performs flagging, primary calibration with 3C 286, phase calibration with J1804+0101 \citep{Perrott2013}. The pipeline exports calibrated fits files which are then imported to \textsc{casa} (version 4.7) for imaging. We cleaned and imaged the data using the \textsc{casa} task \textit{tclean}. 

\subsection{NIR/optical/UV}

\subsubsection{REM}

The NIR photometric monitoring of Swift~J1753 was carried out in H-band ($1.662 \, \rm \mu m$) with the REMIR infrared camera mounted on the REM (Rapid Eye Mount) telescope at La Silla Observatory. Observations span 57 epochs from February 5 to April 19, 2024 (MJD 60345–60419). At each epoch, 4 jitter cycles were executed (NINT = 4), each consisting of 5 dithered exposures of 15 s (DIT = 15 s, NDIT = 5), for a total of 20 frames and a net on-source integration time of 300 s per epoch. Within each jitter cycle, the median combination of the 5 spatially shifted frames yielded an estimate of the sky background, which was then subtracted from each individual frame. The sky-subtracted frames were aligned and median-combined to produce a single nightly average image per epoch. Aperture photometry was subsequently performed on each nightly average using {\tt daophot} \citep{Stetson1987} to extract the H-band flux of the source. Flux calibration was performed against a group of seven field stars with H-band magnitude tabulated in the 2MASS catalog.

\subsubsection{LCO}

Swift~J1753 was monitored with the 1-m and 2-m telescopes of the Las Cumbres Observatory (LCO) global network \citep{2013PASP..125.1031B} as part of an ongoing optical monitoring programme of approximately 50 low-mass X-ray binaries \citep{2008arXiv0811.2336L}. Previous LCO monitoring of Swift~J1753 between 2010 and 2017 has been presented by \citet{2013MNRAS.433..740S} and \citet{2019ApJ...876....5Z}. LCO Observations for this study were obtained between MJD~60200 and MJD~60600 in up to eight optical bands ($B$, $g'$, $V$, $r'$, $R$, $i'$, $z_{\rm s}$, and $Y$), with the filter selection varying between epochs. The raw science images were processed using the standard LCO BANZAI pipeline, which performs bias subtraction, dark-current correction, flat-fielding, and initial astrometric calibration. Photometric measurements were extracted using the XB-NEWS reduction pipeline \citep{Russell2019_XBNEWS, 2026MNRAS.546ag138S} which performs automated quality assessment, image alignment, source detection, photometric calibration against reference catalogues, and extraction of calibrated optical light curves. An astrometric solution was computed for each image using Gaia DR2 positions, and photometric calibration was performed against field stars from the ATLAS-REFCAT2 catalogue \citep{2018ApJ...867..105T}. In cases where the source was not automatically detected above the pipeline detection threshold, forced photometry was performed at the known position of Swift~J1753. Data points with photometric uncertainties greater than 0.25 mag were rejected from the final light curves.

\subsubsection{ATLAS}

Optical photometry of Swift~J1753 was also obtained from the Asteroid Terrestrial-impact Last Alert System \citep[ATLAS;][]{2018PASP..130f4505T}, an all-sky survey operating four 0.5 m telescopes in Hawaii, Chile, and South Africa. Forced photometry at the known position of the source was obtained through the ATLAS forced photometry server \citep{2021TNSAN...7....1S} in the $o$ (orange, 560--820~nm) and $c$ (cyan, 420--650~nm) bands. We used the non-image-subtracted photometry products rather than the default difference-imaging measurements because the prolonged activity of the source may affect the ATLAS reference templates. The ATLAS data obtained are used exclusively in the broad-band SED analysis.

\subsubsection{\textit{Swift}/UVOT}

Swift~J1753 was monitored by the UV/Optical Telescope (UVOT) onboard the Neil Gehrels \textit{Swift} Observatory \citep{Gehrels2004_Swift} in four or five (depending on the epoch) of its six optical and UV filters (\textit{v, b, u, uvw1, uvm2, uvw2}). \textit{Swift}/UVOT data were pre-processed at the Swift Science Data Center and were downloaded from the Swift archive. We then used the \textsc{uvotsource} task to perform aperture photometry, with a circular extraction region of radius $5''$.

\subsection{X-rays}

\subsubsection{\textit{Swift}/XRT}

The X-ray activity of Swift~J1753 was monitored by the Neil Gehrels \textit{Swift} Observatory \citep{Gehrels2004_Swift} with the X-Ray Telescope (XRT), mainly as part of the SwiftKAT follow-up program, for a total of 31 epochs (see Table \ref{tab:obs_swift}). The XRT was operated in Photon Counting (PC) mode during all the epochs, except on MJD 60246 and MJD 60253 when the Windowed Timing (WT) mode was used instead. The data reduction was carried out with the \textsc{heasoft} package (version 6.33.1). We first ran the \textsc{xrtpipeline} task to reprocess the data and create exposure maps. We used the \textsc{xselect} task to extract the source spectra with different extraction regions depending on the observing mode and count rate. A circular region of 20 pixels radius centered on the target position was used in both modes, except when the count rate exceeded $\sim 0.5~\mathrm{counts~s}^{-1}$ in PC mode. In that case, to prevent pile-up from affecting the observation, we extracted the spectrum using an annulus extraction region of variable inner and outer radii. We created ancillary response files with \textsc{xrtmkarf} and used the up-to-date response matrix files from the HEASARC calibration data base.
Spectral analysis was performed with the X-Ray Spectral Fitting Package \citep[\textsc{xspec};][]{xspec}, adopting a simple absorbed power-law model \texttt{tbabs$*$powerlaw}. We first fitted all the spectra while allowing the neutral hydrogen absorption column density ($N_\mathrm{H}$) to vary, and then froze it to the average value of $2.64 \times 10^{21}~\mathrm{cm}^{-2}$, assuming that the same $N_\mathrm{H}$ value applies to all of the spectra. We finally computed the unabsorbed X-ray fluxes in the 1–10 keV energy band using the \texttt{cflux} convolutional model.

\subsubsection{NICER}

The Neutron Star Interior Composition Explorer \citep[NICER;][]{Gendreau2016_NICER} observed Swift~J1753 on 2023 October 6 (MJD 60223; Obs. ID: 6100550108), during the rise of the outburst. We made use of the \textsc{nicerl3-spect} pipeline from the \textsc{heasoft} package (version 6.36, which supports data taken after the light leak event) to generate a spectrum, background and associated responses, considering energies above 0.5~keV for spectral fitting.

\subsubsection{INTEGRAL/IBIS}

The INTEGRAL satellite observed Swift\,J1753 during some part of satellite revolution 2699 (MJD\,60236--60238). We relied on data from the Imager on-Board the INTEGRAL Satellite \citep[IBIS;][]{Ubertini_2003}, in particular its upper detection plane, the IBIS Soft Gamma Ray Imager \citep[ISGRI;][]{Lebrun_2003}. The spectrum was extracted with the INTEGRAL Off-line Scientific Analysis (OSA) software  v11.2, following standard reduction processes\footnote{\url{https://www.isdc.unige.ch/integral/download/osa/doc/11.2/osa_um_ibis/man_html.html}}. Swift\,J1753 was in the FOV of IBIS on MJD\,60236 for an effective exposure of 33.8\,ks. The flux analysis was performed over 18 logarithmically spaced energy bins in the 15--445\,keV range. The bins below 34\,keV were ignored during parameter fitting, since the low energy bins are known to suffer from high systematic error due to the aging of the CdTe detector. Additionally, a systematic error of 2\% was added to all energy channels, as recommended by the user manual. Unfortunately, the source was outside the smaller FOV of the JEM-X instrument \citep{Lund2003}, which prevented us from obtaining data at lower energies than 34\,keV with INTEGRAL.

\subsection{Archival observations}

The initial 11-year-long outburst (2005-2016), as well as the 2017 mini-outbursts, were regularly monitored in radio and X-rays. In this paper, we make use of the fluxes reported in \citet{CadolleBel2007}, \citet{Soleri2010}, \citet{Tomsick2015}, \citet{Rushton2016} and \citet{Plotkin2017}.

\section{Results}
\label{sect:results}

\subsection{Outburst evolution}
\label{subsec:outburst_evolution}

\begin{figure*}
    \centering
    \includegraphics[width=0.95\linewidth]{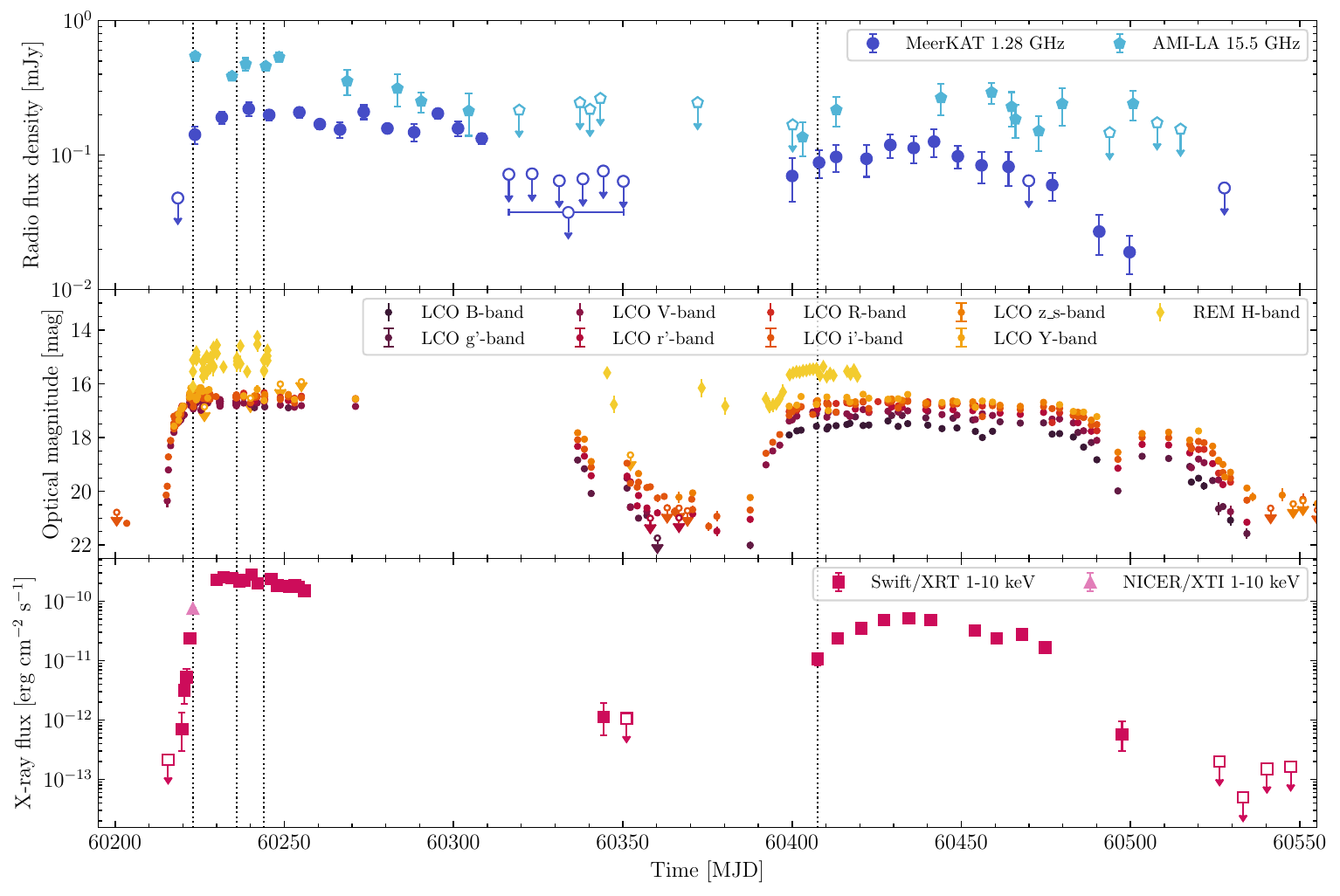}
    \caption{Multiwavelength light curves of the 2023-2024 outburst of Swift~J1753. \textit{Top panel}: MeerKAT 1.28 GHz and AMI--LA 15.5 GHz radio flux density. \textit{Middle panel}: LCO and REM optical/NIR magnitude. \textit{Bottom panel}: \textit{Swift}/XRT and NICER/XTI 1--10 keV unabsorbed X-ray flux. In all panels, the upper limits are marked with empty symbols and downward arrows. The vertical dotted lines mark the epochs chosen for SED fitting.}
    \label{fig:lightcurves}
\end{figure*}

Radio, NIR/optical and X-ray light curves are shown in Fig. \ref{fig:lightcurves}. In the X-rays, \textit{Swift}/XRT detected Swift~J1753 in 29 of the 35 observations. We detected the source ($> 3\sigma$) in 26 of the 35 MeerKAT observations, and 18 of the 27 AMI--LA observations we carried out. When detected, the target appeared as an unresolved point source at the phase center, and no discrete transient ejecta were observed. From the double-hump shape of the light curves, the temporal evolution of the source can clearly be divided into two periods of time: a first period of activity, that spanned $\sim 5$ months, followed by a (slightly weaker) outburst phase, that lasted $\sim 4$ months. A short re-brightening followed, as seen in the optical light curves (MJD 60500--60540), but it was too weak and/or too short to be detected in the X-ray and radio bands. During the first part of the outburst, the dense X-ray monitoring allowed us to follow the rise and the beginning of the decay phases, until the source could not be observed due to Sun constraints. From MJD 60215 to MJD 60232, the source rapidly brightened in soft X-rays, going from $\la 2 \times 10^{-13}~\mathrm{erg~cm^{-2}~s^{-1}}$ to a peak of $\sim 2.7 \times 10^{-10}~\mathrm{erg~cm^{-2}~s^{-1}}$. From MJD 60234 on, Swift~J1753 progressively faded until \textit{Swift} observations were stopped due to Sun constraints. The entirety of the second part of the outburst was monitored with \textit{Swift} on a weekly cadence.

At radio frequencies, the source stayed relatively faint during the two phases of the outburst. After a non-detection at 1.28 GHz on MJD 60218, with a $3\sigma$ (where $\sigma$ is the rms noise level in a source-free region) upper limit of $48~\mu\mathrm{Jy}~\mathrm{bm}^{-1}$, the flux density remained rather stable at the level of $\sim 0.2~\mathrm{mJy}$ until MJD $\sim 60310$. On MJD 60316, the source was not detected, with an upper limit of $72~\mu\mathrm{Jy}~\mathrm{bm}^{-1}$, and we continued to monitor it for five more weeks. We stacked the six January/February MeerKAT observations (MJD 60316--60350) to get a better sensitivity, leading to an upper limit of $37~\mu\mathrm{Jy}~\mathrm{bm}^{-1}$. The upper limits in radio occur during the "dip" between the two humped morphology as seen in X-rays and optical, and are thus consistent with a correlation across the electromagnetic spectrum. During the second phase of the outburst, the radio flux evolved very smoothly between our first observation (MJD 60400) and the last detection on MJD 60499, reaching a low flux density peak of $\sim 120~\mu\mathrm{Jy}~\mathrm{bm}^{-1}$ at 1.28 GHz on MJD 60441. Throughout the outburst, the AMI--LA flux densities were always larger than the MeerKAT flux densities, indicating an inverted radio spectrum.

In order to study the radio spectrum of Swift~J1753, we took advantage of the multi-band radio monitoring with MeerKAT ($\nu_1 = 1.28~\mathrm{GHz}$) and AMI-LA ($\nu_2 = 15.5~\mathrm{GHz}$). To determine a value of the spectral index $\alpha$ ($S_\nu \propto \nu^\alpha$, where $S_\nu$ is the flux density at a given frequency $\nu$) for each MeerKAT epoch during the outburst, we linearly interpolated the AMI light curve using the \textsc{scipy} package and computed $\alpha$ and its uncertainty $\sigma(\alpha)$ as
\begin{equation*}
    \alpha = \frac{\log (S_{\nu_1} / S_{\nu_2})}{\log (\nu_1/\nu_2)}
    ,\;\;
    \sigma(\alpha) = \frac{1}{\ln\left(\nu_1/\nu_2\right)} \sqrt{\left(\frac{\sigma(S_{\nu_1})}{S_{\nu_1}}\right)^2 + \left(\frac{\sigma(S_{\nu_2})}{S_{\nu_2}}\right)^2}
\end{equation*}
where $S_{\nu_1}$, $S_{\nu_2}$, $\sigma(S_{\nu_1})$ and $\sigma(S_{\nu_2})$ are the flux densities and their associated uncertainties at frequencies $\nu_1$ and $\nu_2$.

We report in Table \ref{tab:obs_swift} the final parameters of the spectral fit for each \textit{Swift} epoch. During the whole duration of the double-peaked outburst, the X-ray spectrum was well described by an absorbed power-law model, with a photon index that never exceeded 1.8\footnote{Except at the beginning of the outburst on MJD 60221, when the $\Gamma$ was poorly constrained ($\Gamma = 2.13 \pm 0.37$) due to a low number of counts.} ($\sim1.6$ on average). We therefore conclude that the source remained in the hard state during the whole period of activity, and label the 2023-2024 outburst as "hard-only". This is consistent with the fact that we detect flat-spectrum radio emission continuously during the outburst and that no radio flare nor transient radio ejecta were observed.

\subsection{Constructing a smooth $L_\mathrm{X}$--$L_\mathrm{R}$ plane for Swift~J1753}
\label{subsection:correlation}

To investigate the behavior of Swift~J1753 in the radio/X-ray plane, we paired up MeerKAT and \textit{Swift}/XRT observations to get simultaneous measurements of the radio and X-ray luminosities. We converted the flux density $S_\nu$ at $\nu = 1.28~\mathrm{GHz}$ to the 5--GHz monochromatic radio luminosity $L_\mathrm{R} = 4\pi D^2 \nu S_\nu$ using the spectral indices determined Sect. \ref{subsec:outburst_evolution} and assuming a distance $D = 3.9~\mathrm{kpc}$ \citep{Yanes-Rizo2025}. The \textit{Swift}/XRT 1--10 keV unabsorbed flux $F_\mathrm{X}$ was converted to the X-ray luminosity $L_\mathrm{X} = 4\pi D^2 F_\mathrm{X}$. We obtain a total of 19 measurements of X-ray and radio luminosities, that we report in Table \ref{tab:correlation}.

Previous monitoring campaigns during the 2005-2016 initial outburst and the 2017 mini-outbursts \citep[e.g.,][]{Soleri2010, Rushton2016, Plotkin2017} covered a wide range of X-ray luminosities ($10^{33}$--$10^{37}~\mathrm{erg~s^{-1}}$) in the $L_\mathrm{X}$--$L_\mathrm{R}$ plane, making Swift~J1753 one of the most well-studied outlier sources of the correlation. Yet, at higher luminosities ($L_\mathrm{X} \geq 10^{36}~\mathrm{erg~s^{-1}}$), the radio-quiet track is significantly scattered, as illustrated in the top panel of Fig. \ref{fig:radio_xray_corrected}. After an in-depth look at the measurements from various studies on Swift~J1753, we noticed that:
\begin{itemize}
    \item X-ray luminosities had not always been computed in the same energy band;
    \item there are several duplicate measurements (computed slightly differently);
    \item the 5--GHz radio luminosities were calculated from measurements performed at different frequencies, assuming a flat radio spectrum, an assumption that is not always verified.
\end{itemize}

\begin{figure}
    \centering
    \includegraphics[width=\linewidth]{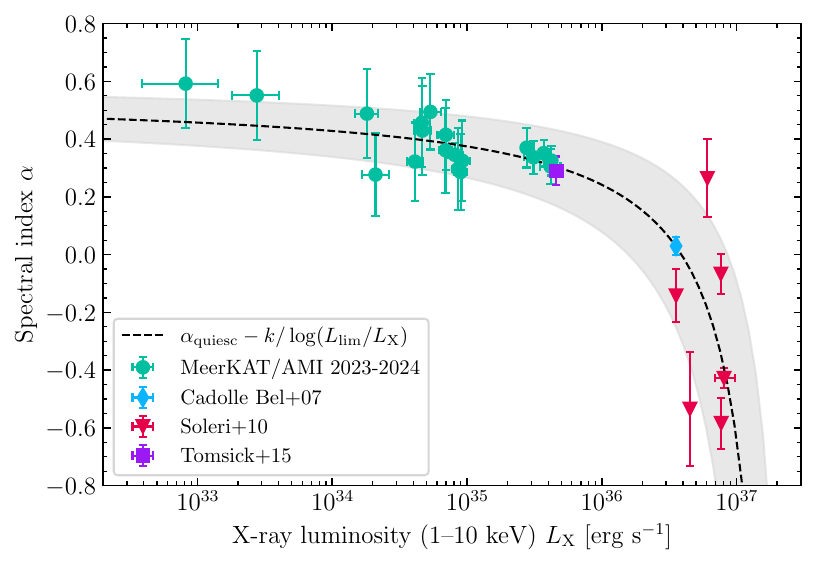}
    \caption{Spectral index versus X-ray luminosity. Green circles correspond to our measurements, while cyan, red and purple symbols mark spectral indices calculated using data from \citet{CadolleBel2007}, \citet{Soleri2010} and \citet{Tomsick2015}, respectively. A tentative fit along with its $1\sigma$ confidence interval are shown in dashed line and shaded area.}
    \label{fig:alpha_vs_Lx}
\end{figure}

\begin{figure*}
    \sidecaption
    \includegraphics[width=12cm]{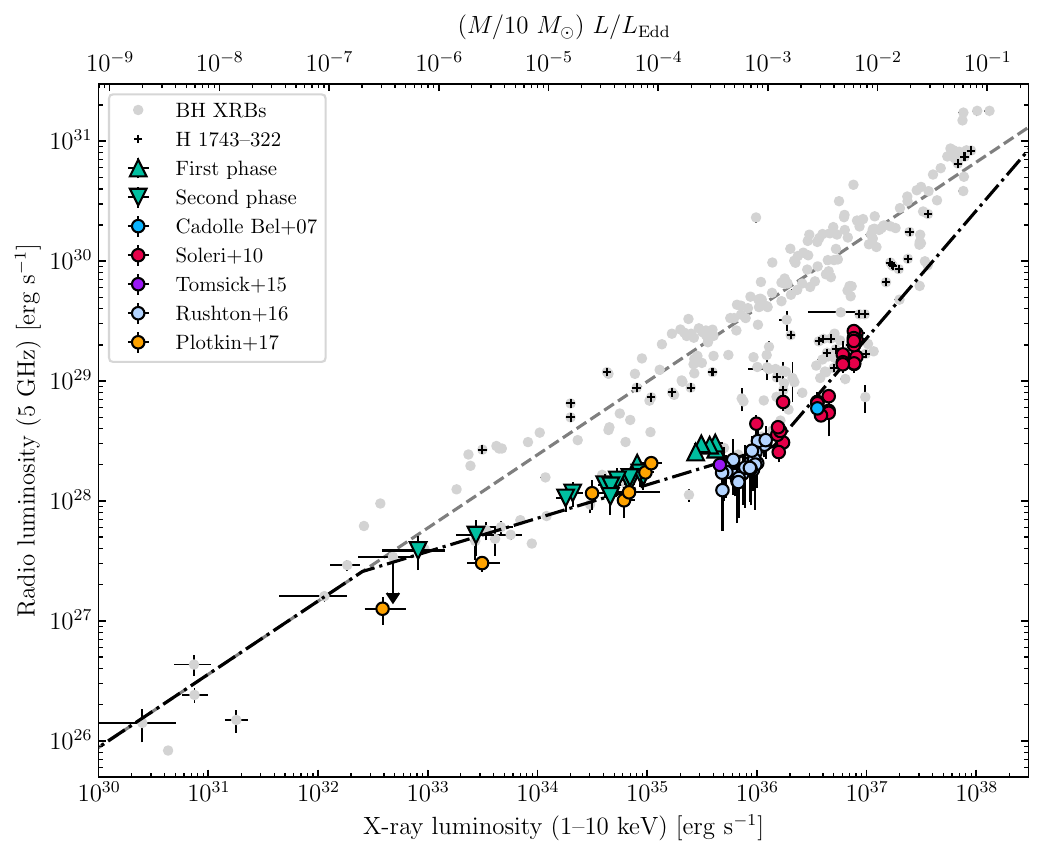}
    \caption{The radio/X-ray luminosity plane of the BH XRB Swift~J1753, assuming it is located 3.9 kpc away \citep{Yanes-Rizo2025}. We include data from this work (green upward and downward triangles for the first and second phases of the 2023-2024 outburst) and previous studies \citep{CadolleBel2007, Soleri2010, Tomsick2015, Rushton2016, Plotkin2017}. Luminosities were calculated following the procedure detailed in Sect. \ref{subsection:correlation}. Downward arrows represent upper limits. For comparison, we show data from H~1743--322 \citep{Coriat2011} as representative of the outlier branch of the correlation and a compilation of BH XRBs from \citet{Bahramian2022}. The standard track ($\beta = 0.61$), marked as the gray dashed line, is shown for illustrative purposes. The black dash-dotted line corresponds to the broken power-law fit described in Sect. \ref{subsection:correlation}, with the following best-fit parameters: $\beta_\mathrm{trans} = 0.28 \pm 0.03$, $\beta_\mathrm{outl} = 1.06\pm 0.11$, $L_\mathrm{trans} = (1.4 \pm 0.6) \times 10^{36} (D/3.9~\mathrm{kpc})^2~\mathrm{erg~s^{-1}}$.}
    \label{fig:correlation}
\end{figure*}

To test the latter hypothesis, we first examined the possibility of a significantly variable spectral index. Indeed, while \citet{CadolleBel2007} determined a value compatible with a flat spectrum ($\alpha = 0.03 \pm 0.03$) in 2005, \citet{Tomsick2015} measured an inverted spectrum ($\alpha = 0.29 \pm 0.05$) in 2014. In Fig. \ref{fig:alpha_vs_Lx}, we plot these measurements, as well as those from our observations and those calculated from flux densities reported by \citet{Soleri2010}, as a function of the X-ray luminosity. The spectral index, which ranges from $-0.6$ to $+0.6$, appears to depend on the X-ray luminosity: for $10^{33}~\mathrm{erg~s^{-1}} \leq L_\mathrm{X} \leq 10^{35-36}~\mathrm{erg~s^{-1}}$, $\alpha$ slowly decreases as $L_\mathrm{X}$ increases. At higher luminosities ($L_\mathrm{X} \geq 10^{36}~\mathrm{erg~s^{-1}}$), the behavior of $\alpha$ is less clear because of the large scatter of the data (mostly from \citealp{Soleri2010}), although it seems to drop much faster as $L_\mathrm{X}$ increases than at lower luminosities. We fitted the spectral index with a tentative empirical model to interpolate $\alpha$ at different X-ray luminosities, using \texttt{curve\_fit} from the \textsc{scipy} package:
\begin{equation*}
    \alpha = \alpha_\mathrm{quiesc} - k / \log (L_\mathrm{lim}/L_\mathrm{X}).
\end{equation*}
Here $\alpha_\mathrm{quiesc}$ is the asymptotic spectral index at very low luminosities (i.e., close to quiescence), $k$ is a normalization constant and $L_\mathrm{lim}$ is a limit luminosity at which the spectral index diverges (i.e., the radio spectrum becomes very steep). We find $\alpha_\mathrm{quiesc} = 0.55 \pm 0.06$, $k = 0.42 \pm 0.11$ and $\log L_\mathrm{lim} = 37.4 \pm 0.1$. Interestingly, \citet{Hughes2025b} found the opposite behavior in Swift J1727.8--1613, i.e., the spectral index appeared to get steeper at lower luminosity hard state.

In order to improve the estimation of luminosities, we applied the following method to the archival and new data:
\begin{enumerate}
    \item Interpolate the X-ray light curve in log space with \textsc{scipy} to maximize simultaneity with radio measurements;
    \item Compute $L_\mathrm{X}$ after converting the unabsorbed fluxes to the 1--10 keV energy band (assuming $\Gamma = 1.6$ when not known);
    \item Compute $L_\mathrm{R}$ at 5 GHz using the spectral index: either a measured value if the observation is multi-band; or an inferred value from the fit shown in Fig. \ref{fig:alpha_vs_Lx};
\end{enumerate}
In Fig. \ref{fig:correlation}, we plot these luminosities (also reported in Table \ref{tab:correlation}), along with data from this work and other BH XRBs, to allow a direct comparison. We also include measurements of H~1743--322 \citep{Coriat2011} as representative of the radio-quiet sources. We note that our measurements comprise detections at $L_\mathrm{X} \la 10^{34}~\mathrm{erg~s^{-1}}$, a region of the $L_\mathrm{X}$--$L_\mathrm{R}$ plane that remains poorly explored, yet essential to probe the accretion/ejection coupling at the lowest accretion rates. For $L_\mathrm{X} \ga 10^{36}~\mathrm{erg~s^{-1}}$, Swift~J1753 seems to follow a steep correlation, as was previously reported by \cite{Soleri2010}. It is worth noting that a number of these high luminosity measurements have quite negative spectral indices, as shown in Fig. \ref{fig:alpha_vs_Lx}. Even though Swift~J1753 was in the hard state at the time of the observations, it is quite possible, as noted by \citet{Soleri2010}, that the radio emission was contaminated by steep-spectra, optically thin discrete ejecta ejected close to the peak of the outburst. In that case, the actual radio emission from the compact jets would be overestimated and that would bias the outlier track. For $10^{33}~\mathrm{erg~s^{-1}} \la L_\mathrm{X} \la 10^{36}~\mathrm{erg~s^{-1}}$, the source adopts a much flatter correlation. Both measurements from the initial outburst (2005-2017) and the 2023-2024  are consistent with the same track at those X-ray luminosities. For $L_\mathrm{X} \la 10^{32-33}~\mathrm{erg~s^{-1}}$, the source seems to intersect the standard track. Hence, Swift~J1753 follows the radio-quiet, outlier track at high luminosities and possibly the radio-loud, standard track at the lowest luminosities, transitioning between the two tracks via a shallow branch.

A similar behavior was observed in H~1743--322 \citep{Coriat2011}, MAXI J1348--630 \citep{Carotenuto2021b} and Swift J1727.8--1613 \citep{Hughes2025b}. In order to compare the correlation indices and the transition luminosities, we tentatively fit data from this work and previous studies, adopting a broken power-law model:

\begin{equation*}
    L_\mathrm{R} \propto
    \left\{
    \begin{array}{ll}
       L_\mathrm{X}^{\beta_\mathrm{trans}} & \mathrm{if~} L_\mathrm{X} \leq L_\mathrm{trans}\\
       L_\mathrm{X}^{\beta_\mathrm{outl}} & \mathrm{if~} L_\mathrm{X} \geq L_\mathrm{trans}
    \end{array}
    \right .
\end{equation*}
The result of the fit is shown in black dash-dotted line in Fig. \ref{fig:correlation}. The fit yields values for correlation slopes: $\beta_\mathrm{trans} = 0.28 \pm 0.03$ for the transition branch, and $\beta_\mathrm{outl} = 1.06\pm 0.11$ for the outlier track, which connect at the transition luminosity $L_\mathrm{trans} = (1.4 \pm 0.6) \times 10^{36} (D/3.9~\mathrm{kpc})^2~\mathrm{erg~s^{-1}}$. We note that this luminosity also marks a transition of behavior of the spectral index, as for luminosities above $L_\mathrm{trans}$, $\alpha$ appears to change significantly (see Fig. \ref{fig:alpha_vs_Lx}). We find an estimation of the luminosity at which Swift~J1753 rejoins the standard track, $L_\mathrm{stand} \sim 3 \times 10^{32}(D/3.9~\mathrm{kpc})^2~\mathrm{erg~s^{-1}}$.

\subsection{Broad-band spectral modeling}
\label{sect:SED_modeling}

\begin{figure*}
    \sidecaption
    \includegraphics[width=12cm]{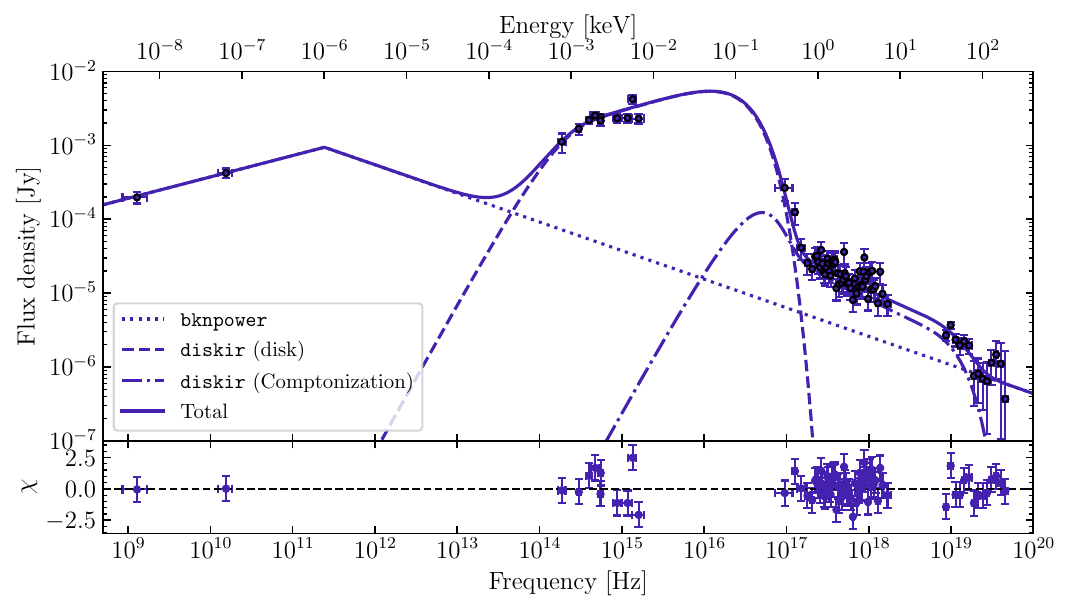}
    \caption{Broad-band spectral energy distribution of Swift~J1753 on MJD 60236. Quasi-simultaneous radio (MeerKAT, AMI--LA), NIR/optical/UV (REM, LCO, \textit{Swift}/UVOT), soft and hard X-ray (\textit{Swift}/XRT, INTEGRAL/IBIS) data are shown. The full multi-wavelength model is \texttt{tbabs$*$(bknpower+diskir)}, and has been fitted with $\Gamma_\mathrm{thin}$ and $kT_e$ left free (\textit{case A} in Table \ref{tab:sed_par} and Sect. \ref{sect:SED_modeling}). Contribution from individual components (disk, Comptonization, broken power-law) are indicated. The X-ray and UV data have been de-absorbed for plotting purposes.}
    \label{fig:sed}
\end{figure*}

\begingroup
\renewcommand{\arraystretch}{1.5}
\begin{table*}
    \caption{Parameters for broad-band SED fits, with the model \texttt{tbabs$*$(bknpower+diskir)}.}
    \centering
    \begin{tabular}{ccccccccccc}
        \hline
        MJD & $N_\mathrm{H}$ & $\Gamma_\mathrm{thick}$ & $\Gamma_\mathrm{thin}$ & $N_\mathrm{bpl}$ & $kT_\mathrm{disk}$ & $\Gamma$ & $kT_e$ & $L_c / L_d$ & $N_\mathrm{disk}$ & $\chi_r^2$ (dof)\\
        & [$10^{21}~\mathrm{cm^{-2}}$] & & & [$\times 10^{2}$] & [ev] & & [keV] & & [$\times 10^{7}$] &\\
        \hline
        60223 & $2.33 \pm 0.16$ & $0.46\tablefootmark{*}$ & $1.7\tablefootmark{*}$ & $61 \pm 11$ & $31_{-15}^{+13}$ & $1.68 \pm 0.04$ & $50\tablefootmark{*}$ & $0.29_{-0.10}^{+0.20}$ & $5.3_{-2.4}^{+6.7}$ & 1.28 (114)\\
        60236 (\textit{A}) & $3.3 \pm 0.9$ & $0.71\tablefootmark{*}$ & $1.39_{-0.03}^{+0.20}$ & $0.78 \pm 0.15$ & $54_{-15}^{+12}$ & $1.63_{-0.09}^{+0.10}$ & $14_{-5}^{+16}$ & $0.26_{-0.09}^{+0.18}$ & $1.2_{-0.5}^{+1.7}$ & 1.08 (76)\\
        60236 (\textit{B}) & $3.7 \pm 0.9$ & $0.71\tablefootmark{*}$ & $> 1.5$ & $0.78 \pm 0.15$ & $55_{-14}^{+13}$ & $1.64_{-0.08}^{+0.12}$ & $50\tablefootmark{*}$ & $0.43_{-0.20}^{+0.09}$ & $1.2_{-0.5}^{+1.3}$ & 1.13 (77)\\
        60236 (\textit{C}) & $3.5 \pm 0.9$ & $0.71\tablefootmark{*}$ & $1.7\tablefootmark{*}$ & $0.78 \pm 0.15$ & $54_{-15}^{+12}$ & $1.61 \pm 0.05$ & $28_{-8}^{+38}$ & $0.41_{-0.11}^{+0.23}$ & $1.3_{-0.5}^{+1.7}$ & 1.10 (77)\\
        60244 & $2.2 \pm 0.4$ & $0.66\tablefootmark{*}$ & $1.7\tablefootmark{*}$ & $1.75 \pm 0.24$ & $40_{-14}^{+8}$ & $1.59 \pm 0.07$ & $50\tablefootmark{*}$ & $0.74_{-0.25}^{+0.77}$ & $2.8_{-1.1}^{+6.0}$ & 0.94 (337)\\
        60407 & $2.64\tablefootmark{*}$ & $0.76\tablefootmark{*}$ & $1.7\tablefootmark{*}$ & $0.12 \pm 0.05$ & $41_{-7}^{+3}$ & $1.61 \pm 0.31$ & $50\tablefootmark{*}$ & $0.041_{-0.022}^{+0.012}$ & $1.9_{-0.4}^{+1.2}$ & 1.36 (22)\\
        \hline
    \end{tabular}
    \tablefoot{We froze $E_\mathrm{break} = 10^{-6}~\mathrm{keV}$, $f_\mathrm{in} = 0.1$, $r_\mathrm{irr} = 1.1$, $f_\mathrm{out} = 0$, $\log (R_\mathrm{out} / R_\mathrm{in}) = 3$, for all epochs. Throughout this table, we quote 90\% confidence errors.
    \tablefoottext{*}{parameter fixed for this fit.}}
    \label{tab:sed_par}
\end{table*}
\endgroup

\begin{figure}
    \centering
    \includegraphics[width=\linewidth]{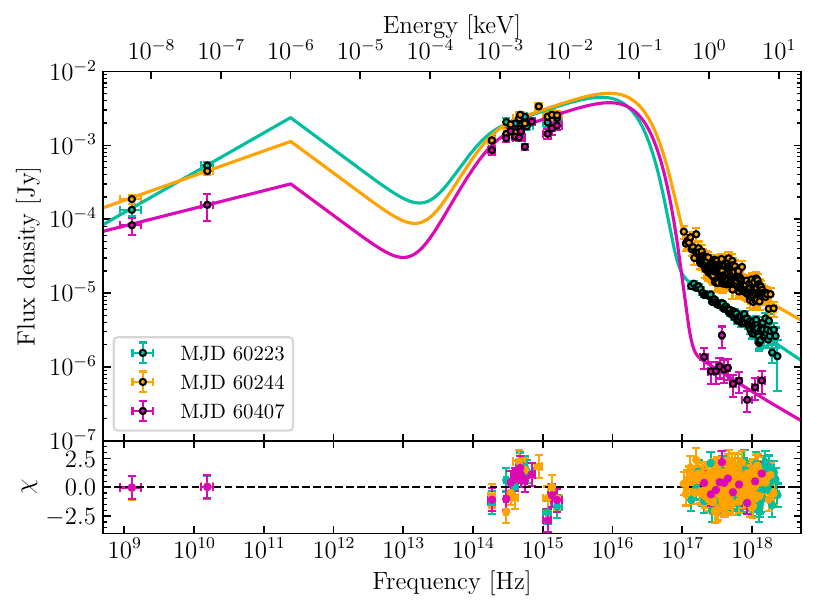}
    \caption{Same as Fig. \ref{fig:sed}, but for three other epochs (MJD 60223, 60244 and 60407).}
    \label{fig:sed_3epochs}
\end{figure}

In the hard state, the compact jets, the accretion disk and the corona all contribute to emission in the NIR/optical/UV and/or X-ray bands. In order to quantify the relative contribution of these components and extract their physical parameters, we performed spectral modeling at several epochs during the outburst. Broad-band spectral energy distributions (SEDs) were built by selecting quasi-simultaneous ($< 1$ day) multi-wavelength data. All broad-band SEDs were modeled with \textsc{xspec} \citep{xspec} version 12.15. We created compatible \textsc{xspec} spectra and response files from the radio and NIR/optical/UV fluxes using the \textsc{ftool} \textsc{flx2xsp}. We obtained four SEDs corresponding to observations on MJD 60223, 60236, 60244 and 60407 (2023 Oct. 6, 19, 27, and 2024 Apr. 7). The epoch on MJD 60236 comprises quasi-simultaneous radio (MeerKAT, AMI--LA), NIR/optical/UV (REM, LCO, \textit{Swift}/UVOT), soft and hard X-ray (\textit{Swift}/XRT, INTEGRAL/IBIS) data and is presented in Fig. \ref{fig:sed}. The three other epochs do not include hard X-ray data but occasionally include additional NICER (for one of them) and ATLAS optical data, and are shown in Fig. \ref{fig:sed_3epochs}.

We used a composite model consisting of a broken power-law (\texttt{bknpower}) and a \texttt{diskir} \citep{diskir} component to fit the full SEDs. While the \texttt{bknpower} component represents the optically thick ($\Gamma_\mathrm{thick}$) and thin ($\Gamma_\mathrm{thin}$) synchrotron emission from the compact jets, the \texttt{diskir} component accounts for both emission from an irradiated accretion disk and Comptonization by the corona. Interstellar grain absorption in X-rays was accounted for with the model \texttt{tbabs} \citep{tbabs}. The IR/optical/UV magnitudes were de-reddened using an extinction of $A_v = 1.395$ \citep[$E(B-V) = 0.45 \pm 0.02$;][]{Rahoui2015}, and considering the relations of \citet{Cardelli1989} and \citet{Mathis1990} to evaluate the absorption coefficients at IR/optical and UV wavelengths, respectively. The full multi-wavelength model is \texttt{tbabs$*$(bknpower+diskir)}. Due to the lack of data in the (sub)-millimeter and far-infrared bands, it is not possible to accurately measure the break energy between the partially self-absorbed and optically thin parts of the synchrotron spectrum (\texttt{bknpower}), we thus fixed its value to $E_\mathrm{break} = 10^{-6}~\mathrm{keV}$ ($\nu_\mathrm{break} \simeq 2.4 \times 10^{11}~\mathrm{Hz}$, i.e., in the millimeter band). The choice of $E_\mathrm{break}$ only weakly affects the inferred $\Gamma_\mathrm{thin}$. We also froze the following $\texttt{diskir}$ parameters, using typical values\footnote{$f_\mathrm{in}$ is the fraction of luminosity in the Compton tail which is thermalized in the inner disk, $f_\mathrm{out}$ is the fraction of bolometric flux thermalized in the outer disk and $R_\mathrm{out}$ (resp. $R_\mathrm{in}$) is the outer (resp. inner) disk radius.}: $f_\mathrm{in} = 0.1$, $r_\mathrm{irr} = 1.1$, $f_\mathrm{out} = 0$, $\log (R_\mathrm{out} / R_\mathrm{in}) = 3$ \citep[e.g.][]{Tomsick2015, Rahoui2015}. The optically thick power-law index $\Gamma_\mathrm{thick}$ only depends on the radio flux measurements and is fixed by the value of the spectral index ($\Gamma_\mathrm{thick} = 1 - \alpha$).

In order to investigate the potential contribution of the compact jets to the X-rays on MJD 60236 (the epoch that included hard X-rays), we considered three cases in our fitting. First, all parameters were left free, except the ones mentioned above (\textit{case A}). We then refitted the data after fixing the electron temperature in the corona to $kT_e = 50~\mathrm{keV}$ \citep[e.g.,][]{Burke2017}, to test if a jet contribution is needed for this typical value of $kT_e$ (\textit{case B}). Finally, we froze the optically thin power-law index $\Gamma_\mathrm{thin}$ to 1.7, to prevent the \texttt{bknpower} component from contributing to the X-ray emission, and let the electron temperature free (\textit{case C}). Due to the absence of hard X-ray observations during the remaining epochs (MJD 60223, 60244 and 60407), we adopted a fitting procedure in which $\Gamma_\mathrm{thin}$ and $kT_e$ are fixed to 1.7 and $50~\mathrm{keV}$, respectively. The broad-band data from each epoch were fit independently. The best-fit parameters are reported in Table \ref{tab:sed_par}.

\section{Discussion}
\label{sect:discussion}

\subsection{A low-luminosity hard-only double-peaked outburst}

Swift~J1753 completed a double-peaked outburst in 2023-2024, that reached soft X-ray unabsorbed fluxes of $\sim 3 \times 10^{-10}~\mathrm{erg~cm^{-2}~s^{-1}}$ (first peak) and $\sim 5 \times 10^{-11}~\mathrm{erg~cm^{-2}~s^{-1}}$ (second peak). Thanks to the low line-of-sight absorption, we were able to detect the source down to $\sim 5 \times 10^{-13}~\mathrm{erg~cm^{-2}~s^{-1}}$. For a $8.8~M_\sun$ black hole at $3.9~\mathrm{kpc}$, these values correspond to a range from $\sim 8 \times 10^{-7}$ to $\sim 5 \times 10^{-4}$ in units of Eddington luminosity $L_\mathrm{Edd}$. As a comparison, the peak flux of the outburst that started in 2005 was $\sim 10^{-8}~\mathrm{erg~cm^{-2}~s^{-1}}$, i.e., $\sim 0.02~L_\mathrm{Edd}$. We thus categorize this outburst as a "low-luminosity" outburst. Likewise, we inferred from the spectral analysis and the persistence of the radio emission (with an inverted spectrum) that the source remained in the hard state during the whole outburst. This overall behavior (hard-only and low-luminosity) is reminiscent of the 2017 mini-outbursts that ended the 12-year-long activity of the source \citep{Plotkin2017, Zhang2019}.

\subsection{A cool, truncated accretion disk}

For each of the four epochs of quasi-simultaneous multiwavelength observations, a multi-temperature disk component successfully reproduces the NIR/optical/UV SED. Throughout the outburst, we find low disk temperatures of $kT_\mathrm{disk} \sim 30$--$55~\mathrm{eV}$. There are hints that the disk marginally contributes to the soft X-ray emission in some of the epochs, but only below $\simeq 0.8~\mathrm{keV}$. In order to test if this cool disk is truncated, we determine its inner radius $R_\mathrm{in}$ from the \texttt{diskir} normalization parameter:
\begin{equation*}
    R_\mathrm{in} / R_g = 0.676 (D/10~\mathrm{kpc}) f^2 \sqrt{N_\mathrm{disk}} \left[ (M_\mathrm{BH} / M_\odot) \sqrt{\cos i}\right]^{-1}
\end{equation*}
where $f$ is the spectral hardening factor \citep[][$f = 1.7$ is a typical value]{ShimuraTakahara1995}. We use the refined parameters found by \citet{Yanes-Rizo2025}: $D = 3.9 \pm 0.7~\mathrm{kpc}$, $M_\mathrm{BH} = 8.8 \pm 1.3 M_\sun$ and $i = 79 \pm 5~\mathrm{deg}$. Following \citet{Tomsick2015}, we assume a low value of $f = 1$ to derive lower limits on the inner radius. From Table \ref{tab:sed_par}, the values of $N_\mathrm{disk}/10^7$ are $5.3_{-2.4}^{+6.7}$, $1.2_{-0.5}^{+1.7}$, $2.8_{-1.1}^{+6.0}$ and $1.9_{-0.4}^{+1.2}$, at MJD 60223, 60236, 60244, 60407, respectively. We obtain $R_\mathrm{in} > 369~R_g$, $> 181~R_g$, $> 282~R_g$ and $> 265~R_g$, implying significant truncation of the disk. The broad-band SED fitting also led \citet{Zhang2010}, \citet{Tomsick2015} and \citet{Rahoui2015} to constrain the disk inner radius to high values of (several) hundreds of gravitational radii. Even after recalculating their $R_\mathrm{in}$ with the same parameter values we used ($D$, $M_\mathrm{BH}$, $i$), we find that $R_\mathrm{in}$ is always measured to be larger than $200~R_g$. These results indicate that the NIR/optical/UV component corresponds to a cool, strongly truncated accretion disk, consistent with a hard spectral state.

It is worth noting that our four epochs correspond to rising phases (MJD 60223 and 60407) and plateau phases (MJD 60236 and 60244) of the X-ray outburst, while for all epochs the optical magnitudes are rather stable, as illustrated in the light curves (Fig. \ref{fig:lightcurves}). The presence of a delay between the optical peak (outer disk) and the X-ray peak (mostly power-law, but also the inner disk) can be interpreted as the propagation of a heating wave through the disk and will be extensively discussed in a forthcoming paper (Russell et al. in prep).

\subsection{Multiwavelength contribution of the compact jets}

Compact jets are known to produce partially self-absorbed synchrotron emission observed as a flat to slightly inverted radio-to-millimeter spectrum ($\alpha \gtrsim 0$), up to a break frequency generally located between the millimeter and near-infrared (NIR) bands \citep[e.g.,][]{CorbelFender2002, Gandhi2011, RussellD2013a, Russell2014}, where the jet plasma becomes optically thin ($\alpha < 0$). The spectral extent of the compact jets above the NIR band is debated. In the model of \citet{MarkoffNowakWilms2005}, the jets can contribute to the X-ray emission via post-shock synchrotron radiation and/or synchrotron self-Compton from the jet base. 

The disk component of the model fits well the NIR/optical/UV data and we do not observe any flux excess at H-band. Thus, we conclude that there is no significant contribution from the compact jets in NIR for the multi-wavelength epochs we considered. However, the H-band light curve has large amplitude variability compared to the optical. REM observations reveal that Swift J1753 was almost $\sim 1~\mathrm{mag}$ brighter during some other epochs, and in that case a NIR excess, consistent with a jet contribution, may be present. The high rms variability of the H-band light curve (particularly during the first part of the outburst, around MJD 60220-60250) is itself consistent with the flickering that compact jets are known to display \citep[e.g.,][]{Casella2010, Gandhi2017, Baglio2018, Vincentelli2018} and this will be further investigated in Russell et al. (in prep).

We use the \textsc{xspec} \texttt{steppar} command to estimate an upper limit on the jet spectral break by computing the evolution of the $\chi^2$ when $E_\mathrm{break}$ is left free. If the jets contribute to the hard X-rays, we find that $E_\mathrm{break} \leq 2 \times 10^{-5}~\mathrm{keV}$ ($\nu_\mathrm{break} \leq 4.8 \times 10^{12}~\mathrm{Hz}$). If not, then a higher value of $\Gamma_\mathrm{thin}$ is possible. When preventing the second part of the broken power-law to be steeper than $\Gamma_\mathrm{thin} = 2$, we derive a less stringent upper limit of $E_\mathrm{break} \leq 1 \times 10^{-4}~\mathrm{keV}$ ($\nu_\mathrm{break} \leq 2.4 \times 10^{13}~\mathrm{Hz}$). During the discovery outburst, \citet{Tomsick2015} found very similar constraints on the break frequency by applying broad-band SED fitting to observations taken in 2014. \citet{Froning2014} came to the same conclusion that the jet does not contribute significantly to the NIR. In contrast, \citet{WangWang2014} detected a mid-IR excess in WISE data, which could originate from a jet or circumbinary disc. \citet{Ulgiati2024} found in X-ray/IR timing observations a possible NIR lag of $\sim 0.7~\mathrm{s}$, which they tentatively interpreted as a long jet lag, hinting a possible jet contribution to the NIR. However, those results are not necessarily inconsistent as the break frequency has been shown to significantly change in several sources \citep[e.g.,][]{Gandhi2011, RussellD2013a, RussellD2013, Russell2014} and the H-band flickering may reflect this.

The location of the spectral break implies that the soft X-rays are not consistent with the extrapolation of the power-law component. Hence, a Comptonization component is strongly required. At higher energies, the hard X-ray emission (possibly the hard tail) might be compatible with the broken power-law component (i.e., the jets), as illustrated in Fig. \ref{fig:sed}. Allowing the powerlaw to contribute to the hard X-rays (i.e., with $\Gamma_\mathrm{thin}$ and $kT_e$ let free, \textit{case A}) yields slightly better fits, but this component is not strongly required, as illustrated by \textit{case B} which gives $\Gamma_\mathrm{thin} > 1.5$ when $kT_e$ is fixed to $50~\mathrm{keV}$. Besides, \textit{case A} leads to a low electron temperature in the corona ($kT_e = 14_{-5}^{+16}~\mathrm{keV}$), which is probably unphysical, although even in the absence of jet contribution ($\Gamma_\mathrm{thin}$ fixed to $1.7$, \textit{case C}), the electron temperature is not necessarily much higher ($kT_e = 28_{-8}^{+38}~\mathrm{keV}$). Overall, the present data do not require a jet contribution unambiguously. Observations in the (sub-)millimeter and mid-IR bands \citep{CorbelFender2002, Russell2014}, as well as polarization measurements in the hard X-rays \citep[e.g.,][]{Laurent2011, Rodriguez2015, Cangemi2023, Bouchet2024}, would be needed to confirm or refute an optically thin synchrotron origin for the hard X-ray component.

It was pointed out by \citet{Soleri2010} that Swift~J1753 is less luminous than expected in the IR/X-ray correlation found by \citet{RussellD2006} (but by a smaller amount that the radio faintness). Besides, the authors noted that its position is reminiscent of the soft-state points of GX 339--4 as reported by \citet{Coriat2009}. In the likely scenario where the NIR emission in the soft state of GX 339--4 is mainly due to the disk, this indicates that Swift~J1753 is both radio- and IR-quiet, i.e., has dim jets. This is in agreement with the results of \citet{RussellD2013a}, who found that radio-quiet BH XRBs have fainter jet breaks (i.e., lower flux at $\nu_\mathrm{break}$) and thus are also IR-faint. We also note that XTE~J1650--500 and H~1743--322, two other radio-quiet sources, have NIR-faint jets \citep{Curran2012, Chaty2015}.

\subsection{The hybrid radio/X-ray correlation of Swift~J1753}

While the high luminosities of the $L_\mathrm{X}$--$L_\mathrm{R}$ plane have been extensively sampled between 2005 and 2016 \citep{Soleri2010, Rushton2016} in the case of Swift~J1753, the region where $L_\mathrm{X} \lesssim 10^{35}~\mathrm{erg~s^{-1}}$ was only explored during the 2017 mini-outbursts of the source \citep{Plotkin2017}. This provided evidence that Swift~J1753 rejoins (or at least approaches) the standard, radio-loud, correlation at lower luminosities. However, since the initial decay was not monitored in the radio and X-rays in 2016, it was unclear whether the source had moved horizontally across the plane, or had faded following the outlier branch down to quiescence and then rose along the standard track during the mini-outbursts. Our observations of the 2023-2024 outburst allow to investigate the transition between the two tracks. During the X-ray rise of the first outburst, the radio flux density remains approximately constant, i.e., the source was moving horizontally across the $L_\mathrm{X}$--$L_\mathrm{R}$ plane. This constitutes a direct observation of the transition between the radio-faint and the radio-loud tracks along a shallow branch. During the second outburst, the peak flux was too low to complete the full transition towards the outlier branch. However, the rise and the decay were slower, thus improving the sampling of the (partial) transition, which is observed back and forth (i.e., during the rise and the decay).

\subsection{Comparison with other outlier/hybrid sources}
\label{subsection:correlation_other_outliers}

\begin{figure}
    \centering
    \includegraphics[width=\linewidth]{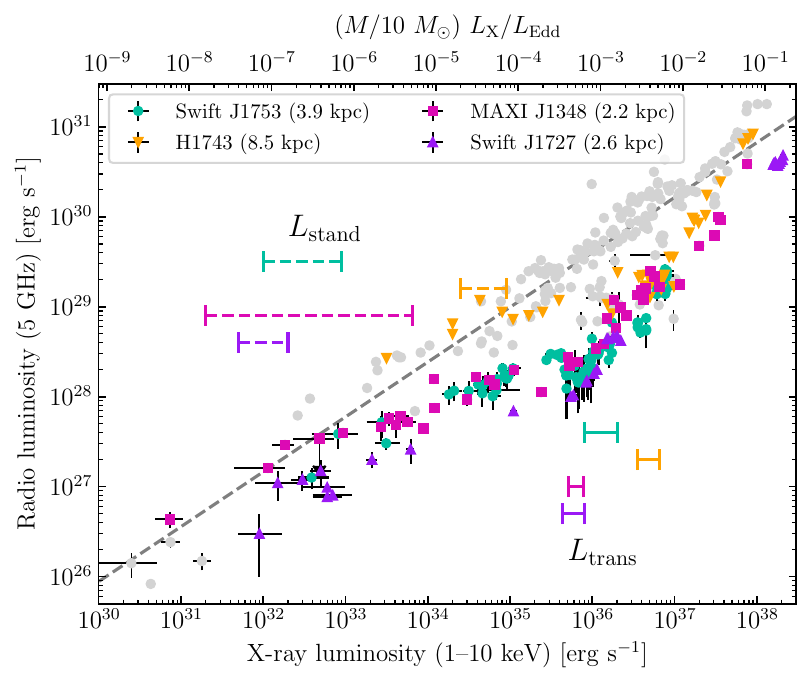}
    \caption{The radio/X-ray luminosity plane of the four best sampled outliers, H1743 \citep{Coriat2011}, MAXI J1348 \citep{Carotenuto2021b, Carotenuto2022b}, Swift J1727 \citep{Hughes2025b} and Swift~J1753 (same data as in Fig. \ref{fig:correlation}). We assume the distances used in the associated references. Horizontal bars indicate the confidence intervals of the transition luminosity $L_\mathrm{trans}$ between the outlier and transition branches (solid line), and the luminosity $L_\mathrm{stand}$ at which the transition branch rejoins the standard branch (dashed line), as determined from a broken power-law fit.}
    \label{fig:correlation_outl}
\end{figure}

Swift~J1753 belongs to a growing class of BH XRBs that display a radio-quiet correlation, including XTE J1650--500 \citep{Corbel2004}, H~1743--322 \citep{Coriat2011, Williams2020}, XTE J1752--223 \citep{Ratti2012, Brocksopp2013}, MAXI J1659--152 \citep{Jonker2012}, MAXI J1535--571 \citep{Russell2019, Parikh2019}, GRS 1739--278 \citep{Xie2020}, MAXI J1631--472 \citep{Monageng2021}, MAXI J1348--630 \citep{Carotenuto2021b, Carotenuto2022b} and Swift J1727.8--1613 \citep{Hughes2025b}. Among these sources, only four have been unambiguously monitored along both the steep and shallow branches: H~1743--322, MAXI J1348--630, Swift J1727.8--1613 (hereafter H1743, MAXI~J1348 and Swift~J1727), along with Swift~J1753. In Fig. \ref{fig:correlation_outl}, we present these sources in the radio/X-ray plane, indicating the transition luminosities $L_\mathrm{trans}$ and $L_\mathrm{stand}$. Here, we have used $D = 2.2_{-0.6}^{+0.5}~\mathrm{kpc}$ for MAXI~J1348 \citep{Chauhan2021}, $D = 2.6_{-1.1}^{+2.7}~\mathrm{kpc}$ for Swift~J1727 \citep{Hughes2025b} and $D = 3.9 \pm 0.7 ~\mathrm{kpc}$ for Swift~J1753 \citep{Yanes-Rizo2025}. Interestingly, the latter outliers occupy a similar parameter space in the $L_\mathrm{X}$--$L_\mathrm{R}$ plane. In particular, these sources have comparable normalizations, correlation slopes ($\beta_\mathrm{trans}\sim 0.2$--$0.3$, $\beta_\mathrm{outl}\sim 1$) and transition luminosities $L_\mathrm{trans} \sim 10^{36}~\mathrm{erg~s^{-1}}$ (i.e., $\sim 10^{-3}~L_\mathrm{Edd}$ for a $10~M_\sun$ black hole). On the other hand, H1743 has a steeper outlier branch ($\beta_\mathrm{outl} = 1.38 \pm 0.03$) than the other main hybrid sources, MAXI J1348 ($0.95 \pm 0.04$), Swift J1727 ($1.02 \pm 0.01$) and Swift~J1753 ($1.06 \pm 0.11$) \citep{Coriat2011, Carotenuto2021b, Hughes2025b}. A caveat is that the distance of Swift~J1727 could be as high as $5.5_{-1.1}^{+1.4}~\mathrm{kpc}$ \citep{Burridge2025}, but in that case its track in the $L_\mathrm{X}$--$L_\mathrm{R}$ plane would be even more consistent with Swift~J1753 and MAXI~J1348. Likewise, there is a tension on the distance of Swift J1753 between the value of \citet{Yanes-Rizo2025} and previous parallax measurements more consistent with $D \simeq 8~\mathrm{kpc}$ \citep{Gandhi2019, Atri2019}, and we conservatively show the radio/X-ray plane for those two distances in figure \ref{fig:radio_xray_distance}.

\subsection{Spectral index vs X-ray luminosity}

When investigating the evolution of the spectral index, we noticed the clear dependence (at least below $10^{36}~\mathrm{erg~s^{-1}}$) of $\alpha$ with respect to the X-ray luminosity $L_\mathrm{X}$, as illustrated in Fig. \ref{fig:alpha_vs_Lx}. We tentatively modeled the evolution of $\alpha$ with an empirical relation, $\alpha = \alpha_\mathrm{quiesc} - k / \log (L_\mathrm{lim}/L_\mathrm{X})$, where the asymptotic, maximum value of $\alpha$ is reached in quiescence ($\alpha_\mathrm{quiesc}$) and $\alpha$ slowly decreases with $L_\mathrm{X}$, up to a luminosity $L_\mathrm{lim}$ at which it drops quickly. Of course, this part of the plot \citep[above $10^{36}~\mathrm{erg~s^{-1}}$, from][]{Soleri2010} is rather scattered and the real behavior at high luminosities remains unclear. As discussed above, some spectral indices are significantly negative and contamination from discrete ejecta cannot be excluded. Another possibility involves the shift of the break frequency towards low frequencies as the quenching of the compact jets begins. In that case, the radio spectral index would probe the optically thin part of the synchrotron spectrum and is expected to be negative.

We note that \citet{EspinasseFender2018} showed that radio-loud sources have on average inverted radio spectra, while radio-quiet sources tend to have slightly negative spectral indices. Swift~J1753, as a radio-quiet source, does not appear to follow this behavior, except at the highest luminosities, at which $\alpha \la 0$. Interestingly, the turnover between the region where $\alpha$ slowly decreases, and the drop at high luminosities, is around $10^{36}~\mathrm{erg~s^{-1}}$, which coincides with the transition luminosity $L_\mathrm{trans}$ between the shallow and the steep branches, hinting a possible connection between these two changes of regime.

\citet{KoljonenRussell2019} suggested that radio-loud sources could have self-absorbed jets with continuous energy dissipation occurring via internal shocks, and radio-quiet sources could result from discrete dissipation by shocked ejecta within the jet. In that case, \citet{KoljonenRussell2019} pointed out that the jet morphology should change near the transition to/from the steep outlier track, resulting in a shift in spectral index, which is essentially what we observe. As mentioned above, \citet{Hughes2025b} reported a similar evolution of the spectral index during the decaying hard state of Swift~J1727, but in the opposite direction (i.e., more optically thin at lower luminosities).

We make an attempt to interpret the evolution of the spectral index in the framework of self-absorbed jet models developed by \citet{Kaiser2006}, who extended previous standard models of jets that did not include energy losses or non-conical geometries \citep[e.g.,][]{BlandfordKonigl1979, HjellmingJohnston1988}. The author considers two classes of jet models: ballistic (conical) jets and adiabatic jets. We do not examine the case of ballistic jet models as they lead to constant spectral indices (for a given electron distribution index), and restrict our discussion to adiabatic jets. The latter can have a non-conical shape whose radius $r(x)$ along the jet axis $x$ is given by $r(x) \propto x^a$, where $a$ parametrizes the confinement of the jet ($0 \leq a \leq 1$ and $a = 1$ corresponds to the conical geometry). \citet{Kaiser2006} formulated three adiabatic jet models, depending on the configuration of the magnetic field: perpendicular to the jet axis, parallel to it and isotropic. In absence of energy losses, the spectral index of the self-absorbed synchrotron spectrum is given by $\alpha = [5 + (p+4)a_5]/2$, where $a_5$ is a function of $a$ and the electron energy distribution index $p$, and its expression is provided for each model in Table 1 from \citet{Kaiser2006}. When adiabatic and radiative energy losses are taken into account, the spectral index is instead $\alpha = (7a + 6a_2 - 3)/(4a + 3a_2)$. Here, $a_2 = a$ for a perpendicular magnetic field, $a_2 = 2a$ for a parallel field and $a_2 = 4a/3$ for an isotropic field.

\begin{figure}
    \centering
    \includegraphics[width=1\linewidth]{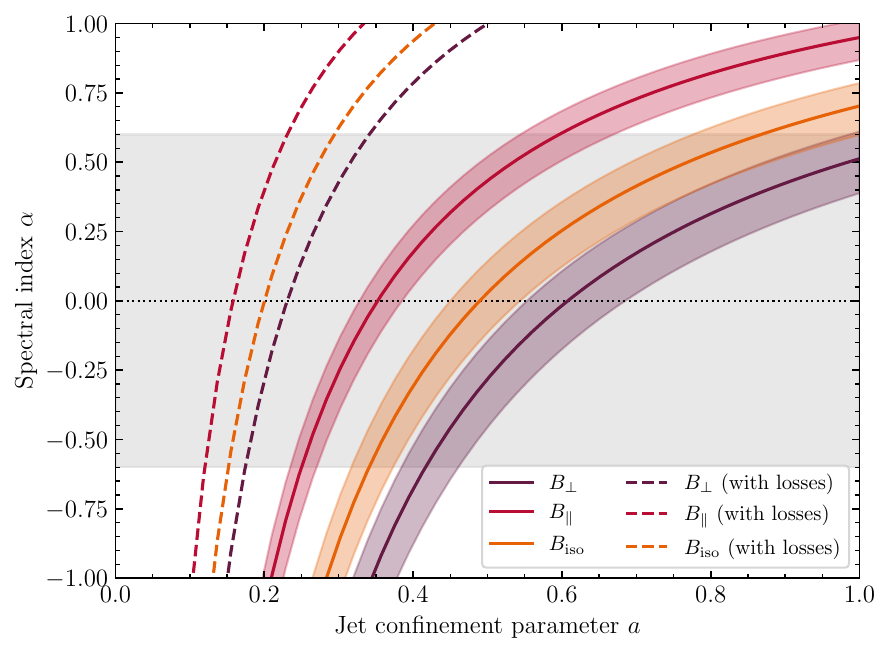}
    \caption{Spectral index versus jet confinement parameter $a$ ($r(x) \propto x^a$) in the models of \citet{Kaiser2006}, for perpendicular ($B_\perp$), parallel ($B_\|$) and isotropic ($B_\mathrm{iso}$) magnetic fields, without (solid line) and with (dashed line) energy losses. For the case without energy losses, $\alpha$ is computed for $p = 2.2$, and shaded regions show the range of $\alpha$ for $1.8 \leq p \leq 2.6$. The horizontal dotted line indicates $\alpha = 0$, while the range of observed spectral indices is represented by the gray shaded area.}
    \label{fig:alpha_confinement}
\end{figure}

\begin{figure}
    \centering
    \includegraphics[width=\linewidth]{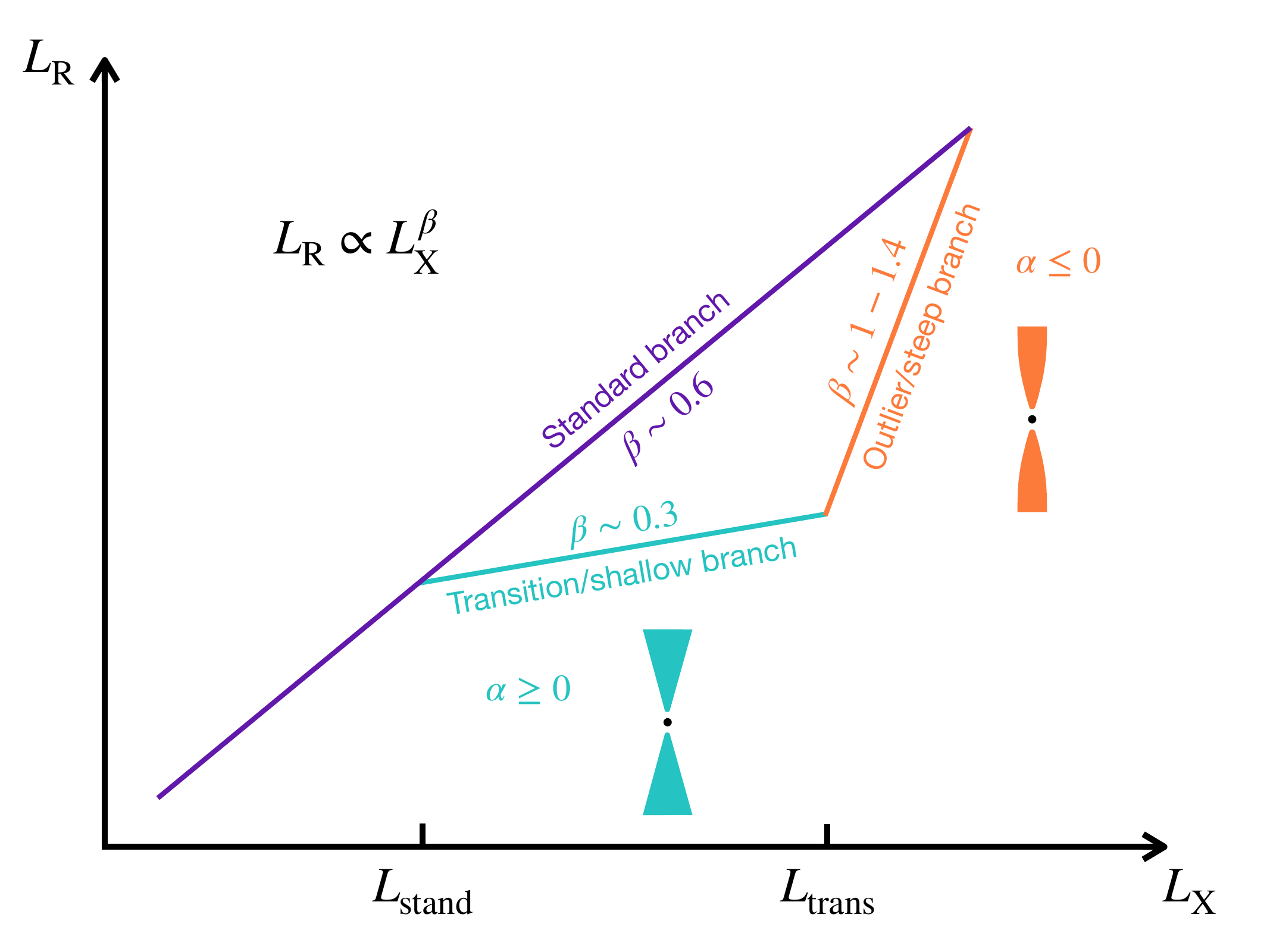}
    \caption{Schematic representation of the radio/X-ray correlation for BH XRBs, along with a tentative scenario for the spectral behavior of Swift~J1753. At high luminosities, outliers follow a steep ($\beta \gtrsim 1$) branch, while below $L_\mathrm{trans}$ they adopt a shallow ($\beta \sim 0.3$), transition branch, that rejoins the standard track ($\beta \sim 0.6$) at $L_\mathrm{stand}$. In Swift~J1753, $\alpha \geq 0$ along the transition branch and $\alpha \leq 0$ at higher luminosities, which in our scenario may corresponds to an evolution of the jet geometry towards stronger confinement as the the accretion rate increases.}
    \label{fig:sketch_correlation_confinement}
\end{figure}

In Fig. \ref{fig:alpha_confinement}, we plot the evolution of the spectral index as a function of the jet confinement parameter $a$, with and without energy losses, for ease of comparison. For all magnetic field configurations and energetic scenarios, $\alpha$ displays a monotonic increase as a function of $a$. The gray shaded area represents the range of observed spectral indices ($-0.6 \la \alpha \la + 0.6$). In principle, all adiabatic jet models we consider are able to cover this range of values. We note that when energy losses are taken into account, much higher confinement (low $a$ value) is required, and parallel magnetic fields imply lower values of $a$ than isotropic and perpendicular fields, the latter being the least constraining (see Fig. \ref{fig:alpha_confinement}). Although $\alpha$ depends on $p$ and $a$ when energy losses are neglected, Fig. \ref{fig:alpha_confinement} clearly shows that a variation of distribution index $p$ cannot account for the range of spectral indices alone. Instead, $\alpha$ has a much stronger dependence on $a$ (i.e., the confinement of the jets) and the evolution of the latter may explain the range of spectral indices. We therefore suggest that the decrease of $\alpha$ for increasing X-ray luminosities is due to a decrease of parameter $a$. In other words, as the accretion rate rises, the jets adopt a more confined geometry, as summarized in Fig. \ref{fig:sketch_correlation_confinement}.

\section{Conclusions}
\label{sect:conclusions}

We presented a multiwavelength study of the 2023-2024 activity of the BH XRB Swift~J1753, based on radio (MeerKAT, AMI--LA), NIR/optical/UV (REM, LCO, \textit{Swift}/UVOT) and X-ray (\textit{Swift}/XRT, NICER, INTEGRAL) observations. The source underwent two consecutive phases of outburst separated by only a short return to quiescence, providing an opportunity to investigate the accretion/ejection coupling over a wide luminosity range. Our main results can be summarized as follows.

The 2023-2024 event consisted of a double-peaked outburst with durations of $\sim5$ and $\sim4$ months, followed by a weak re-brightening detected in the optical. Both episodes remained at low luminosity ($\la 5 \times 10^{-4}~L_\mathrm{Edd}$) and showed persistent hard-state characteristics (absorbed power-laws in X-rays with $\Gamma \lesssim 1.8$). Radio emission lingered throughout the outburst and no transient ejecta were detected. We therefore classify the event as a low-luminosity, hard-only outburst.

Broad-band SED modeling shows that the NIR/optical/UV emission is dominated by a cool accretion disk with temperatures of $kT_\mathrm{disk}\sim25$--$55$ eV. The inferred inner radii ($R_\mathrm{in} \gtrsim 200~R_g$) indicate substantial truncation throughout the outburst, consistent with previous studies and with the persistent hard-state nature of the source. The jet spectral break is constrained to lie below the NIR regime, likely between the radio and mid-IR bands, and we discuss the compact jets contribution in the NIR band. While an optically thin jet contribution to the hard X-rays cannot be excluded, the present data do not require it unambiguously.

Combining the new monitoring with archival data, we revisited the radio/X-ray correlation of Swift~J1753 over $\ga4$ decades in X-ray luminosity. After accounting for the evolution of the radio spectral index, converting measurements to a common energy band and eliminating duplicate measurements, we build a smoother, less scattered radio/X-ray plane for this source. We confirm that Swift~J1753 follows a hybrid track: a steep radio-quiet branch at high luminosities ($\beta_\mathrm{outl} = 1.06 \pm 0.11$), a shallow transition branch ($\beta_\mathrm{trans} = 0.28 \pm 0.03$), and a convergence toward the standard radio-loud track at low luminosities. The transition occurs around $L_\mathrm{trans} \sim 1.4 \times 10^{36}~\mathrm{erg~s^{-1}}$.

We find evidence that the radio spectral index evolves with X-ray luminosity, decreasing from inverted values near quiescence to flatter or negative values at higher luminosities. The turnover appears close to the luminosity where the source leaves the shallow branch and enters the outlier track, suggesting a possible link between jet spectral evolution and changes in the accretion/ejection coupling. Within the framework of adiabatic jet models, this behavior may reflect an evolution of the jet geometry towards stronger confinement at higher accretion rates.

\begin{acknowledgements}

We thank the referee for their careful reading and fruitful comments. The MeerKAT telescope is operated by the South African Radio Astronomy Observatory, which is a facility of the National Research Foundation, an agency of the Department of Science, Technology and Innovation. We acknowledge the use of the ilifu cloud computing facility - \url{www.ilifu.ac.za}, a partnership between the University of Cape Town, the University of the Western Cape, Stellenbosch University, Sol Plaatje University, the Cape Peninsula University of Technology and the South African Radio Astronomy Observatory. The ilifu facility is supported by contributions from the Inter-University Institute for Data Intensive Astronomy (IDIA - a partnership between the University of Cape Town, the University of Pretoria and the University of the Western Cape), the Computational Biology division at UCT and the Data Intensive Research Initiative of South Africa (DIRISA). This work made use of the CARTA (Cube Analysis and Rendering Tool for Astronomy) software (DOI 10.5281/zenodo.3377984 – \url{https://cartavis.github.io}). This work makes use of observations from the Las Cumbres Observatory global telescope network. The authors would like to thank L. Tremou, A. Hughes, F. Carotenuto and P. Saikia for scheduling the MeerKAT observations. MCB acknowledges support from the INAF Fundamental Research Grant XBOOM. We thank Dan Bramich for the creation and implementation of the XB-NEWS pipeline. DMR is supported by Tamkeen under the NYU Abu Dhabi Research Institute grant CASS. TB acknowledges support from the DFG/LIS project SI 2502/6-1, project number 551127478.

\end{acknowledgements}

\bibliographystyle{aa}
\bibliography{bib}

\onecolumn
\begin{appendix}

\section{Radio and X-ray observations}

\begin{table*}[ht!]
    \caption{Summary of MeerKAT radio observations of Swift~J1753.}
    \centering
    \begin{tabular}{lccccc}
        \hline
        Date & Start & Exposure & Robust & RMS\tablefootmark{b} & Flux density\tablefootmark{c}\\
         & MJD & [min] & parameter\tablefootmark{a} & [$\mu\mathrm{Jy~bm^{-1}}$] & [$\mu\mathrm{Jy}$] \\
         \hline
2023-10-01 & 60218.55 & 15 & 0.2 & 16 & $< 48$ \\
2023-10-06 & 60223.61 & 15 & -0.3 & 20 & $142 \pm 22$ \\
2023-10-14 & 60231.54 & 15 & -0.3 & 23 & $191 \pm 20$ \\
2023-10-22 & 60239.49 & 15 & -0.3 & 20 & $221 \pm 25$ \\
2023-10-28 & 60245.53 & 15 & -0.3 & 19 & $199 \pm 18$ \\
2023-11-06 & 60254.44 & 15 & -0.3 & 25 & $208 \pm 18$ \\
2023-11-12 & 60260.49 & 15 & -0.3 & 20 & $170 \pm 14$ \\
2023-11-18 & 60266.48 & 15 & -0.3 & 20 & $155 \pm 20$ \\
2023-11-25 & 60273.47 & 15 & -0.3 & 24 & $210 \pm 25$ \\
2023-12-02 & 60280.46 & 15 & -0.3 & 20 & $158 \pm 8$ \\
2023-12-10 & 60288.35 & 15 & -0.3 & 21 & $148 \pm 22$ \\
2023-12-17 & 60295.37 & 15 & -0.3 & 23 & $204 \pm 17$ \\
2023-12-23 & 60301.33 & 15 & -0.3 & 23 & $158 \pm 20$ \\
2023-12-30 & 60308.33 & 15 & -0.3 & 22 & $133 \pm 12$ \\
2024-01-07 & 60316.34 & 15 & -0.3 & 24 & $< 72$ \\
2024-01-14 & 60323.21 & 15 & -0.3 & 24 & $< 73$ \\
2024-01-22 & 60331.24 & 15 & -0.3 & 21 & $< 64$ \\
2024-01-29 & 60338.22 & 15 & -0.3 & 22 & $< 67$ \\
2024-02-04 & 60344.24 & 15 & -0.3 & 25 & $< 76$ \\
2024-02-10 & 60350.16 & 15 & -0.3 & 21 & $< 64$ \\
2024-01/02\tablefootmark{*} & 60333.90 & $6 \times 15$ & -0.3 & 12 & $< 37$ \\
2024-03-31 & 60400.04 & 15 & 0.2 & 17 & $70 \pm 25$ \\
2024-04-07 & 60407.98 & 15 & 0.2 & 18 & $88 \pm 21$ \\
2024-04-13 & 60413.03 & 15 & -0.3 & 21 & $97 \pm 22$ \\
2024-04-21 & 60421.99 & 15 & -0.3 & 22 & $94 \pm 25$ \\
2024-04-29 & 60429.04 & 15 & -0.3 & 21 & $119 \pm 24$ \\
2024-05-05 & 60435.94 & 15 & -0.3 & 21 & $113 \pm 26$ \\
2024-05-11 & 60441.96 & 15 & -0.3 & 25 & $126 \pm 30$ \\
2024-05-18 & 60448.98 & 15 & -0.3 & 19 & $98 \pm 19$ \\
2024-05-25 & 60455.98 & 15 & -0.3 & 20 & $84 \pm 22$ \\
2024-06-02 & 60463.95 & 15 & -0.3 & 21 & $82 \pm 23$ \\
2024-06-08 & 60469.93 & 15 & -0.3 & 21 & $< 64$ \\
2024-06-15 & 60476.86 & 60 & -0.3 & 12 & $60 \pm 14$ \\
2024-06-29 & 60490.76 & 60 & 0.2 & 10 & $27 \pm 9$ \\
2024-07-08 & 60499.73 & 120 & -0.3 & 7 & $19 \pm 6$ \\
2024-08-05 & 60527.73 & 15 & -0.3 & 19 & $< 57$ \\
    \end{tabular}
    \tablefoot{
    \tablefoottext{*}{Stacking of the six January/February observations.}
    \tablefoottext{a}{Robust parameter used in Briggs weighting scheme.}
    \tablefoottext{b}{RMS noise level estimated from a nearby source free region.}
    \tablefoottext{c}{Radio flux density measured at $1.28~\mathrm{GHz}$ (uncertainties are given at the $1\sigma$ level) if detected, otherwise $3\sigma$ upper limit.}
    }
    \label{tab:obs_meerkat}
\end{table*}

\clearpage
\renewcommand{\arraystretch}{1.3}
\begin{table*}
    \caption{Summary of \textit{Swift}/XRT X-ray observations of Swift~J1753.}
    \centering
    \begin{tabular}{ccccccc}
        \hline
        Date & Start & Obs. ID & Mode\tablefootmark{a} & Exposure & Photon index & 1--10 keV flux\tablefootmark{b}\\
         & MJD & & & [ks] & $\Gamma$ & [$10^{-11}~\mathrm{erg~cm^{-2}~s}$]\\
         \hline
2023-09-28 & 60215.67 & 00016259001 & PC & 1.94 & -- & $< 0.02$ \\
2023-10-02 & 60219.75 & 00016259002 & PC & 0.92 & $1.55$ & $0.07_{-0.04}^{+0.06}$ \\
2023-10-03 & 60220.48 & 00016259003 & PC & 0.97 & $1.52_{-0.61}^{+0.58}$ & $0.32_{-0.13}^{+0.25}$ \\
2023-10-04 & 60221.14 & 00016259004 & PC & 0.97 & $2.13_{-0.37}^{+0.36}$ & $0.53_{-0.14}^{+0.20}$ \\
2023-10-05 & 60222.14 & 00016259005 & PC & 0.89 & $1.67_{-0.16}^{+0.16}$ & $2.35_{-0.32}^{+0.36}$ \\
2023-10-13 & 60230.00 & 00016259006 & PC & 1.09 & $1.54_{-0.15}^{+0.15}$ & $22.95_{-2.91}^{+3.25}$ \\
2023-10-15 & 60232.12 & 00016259007 & PC & 1.16 & $1.59_{-0.14}^{+0.15}$ & $25.52_{-3.17}^{+3.52}$ \\
2023-10-17 & 60234.48 & 00016259008 & PC & 0.92 & $1.55_{-0.16}^{+0.16}$ & $24.37_{-3.27}^{+3.69}$ \\
2023-10-19 & 60236.67 & 00016259009 & PC & 0.71 & $1.44_{-0.21}^{+0.21}$ & $21.38_{-3.67}^{+4.27}$ \\
2023-10-21 & 60238.32 & 00016259010 & PC & 0.83 & $1.63_{-0.17}^{+0.17}$ & $22.53_{-3.21}^{+3.63}$ \\
2023-10-23 & 60240.29 & 00016259011 & PC & 0.82 & $1.39_{-0.17}^{+0.17}$ & $27.92_{-4.02}^{+4.62}$ \\
2023-10-25 & 60242.15 & 00016259012 & PC & 1.06 & $1.66_{-0.15}^{+0.15}$ & $19.93_{-2.54}^{+2.83}$ \\
2023-10-29 & 60246.10 & 00016259013 & WT & 1.32 & $1.65_{-0.04}^{+0.04}$ & $23.76_{-0.80}^{+0.82}$ \\
2023-10-31 & 60248.01 & 00016259014 & PC & 1.24 & $1.53_{-0.14}^{+0.14}$ & $18.29_{-2.05}^{+2.29}$ \\
2023-11-03 & 60251.38 & 00016259015 & PC & 0.71 & $1.55_{-0.20}^{+0.20}$ & $17.86_{-2.86}^{+3.35}$ \\
2023-11-05 & 60253.03 & 00016259016 & WT & 0.49 & $1.54_{-0.08}^{+0.08}$ & $18.43_{-1.18}^{+1.25}$ \\
2023-11-06 & 60254.29 & 00016259017 & PC & 0.82 & $1.37_{-0.19}^{+0.18}$ & $17.59_{-2.64}^{+3.08}$ \\
2023-11-08 & 60256.01 & 00016259018 & PC & 0.84 & $1.44_{-0.17}^{+0.17}$ & $15.11_{-2.09}^{+2.40}$ \\
2024-02-04 & 60344.32 & 00016259019 & PC & 0.82 & $1.55$ & $0.11_{-0.06}^{+0.08}$ \\
2024-02-11 & 60351.11 & 00016259020 & PC & 1.00 & -- & $< 0.11$ \\
2024-04-07 & 60407.59 & 00016259021 & PC & 0.91 & $1.68_{-0.27}^{+0.27}$ & $1.07_{-0.23}^{+0.28}$ \\
2024-04-13 & 60413.50 & 00016259023 & PC & 0.92 & $1.73_{-0.15}^{+0.15}$ & $2.36_{-0.30}^{+0.34}$ \\
2024-04-20 & 60420.41 & 00016259024 & PC & 0.85 & $1.67_{-0.13}^{+0.13}$ & $3.49_{-0.39}^{+0.44}$ \\
2024-04-27 & 60427.19 & 00016259025 & PC & 1.19 & $1.59_{-0.11}^{+0.11}$ & $4.86_{-0.45}^{+0.49}$ \\
2024-05-04 & 60434.58 & 00016259026 & PC & 0.76 & $1.83_{-0.16}^{+0.16}$ & $5.13_{-0.67}^{+0.77}$ \\
2024-05-11 & 60441.01 & 00016259027 & PC & 0.99 & $1.52_{-0.11}^{+0.11}$ & $4.83_{-0.45}^{+0.49}$ \\
2024-05-23 & 60453.99 & 00016259028 & PC & 0.47 & $1.68_{-0.21}^{+0.21}$ & $3.20_{-0.54}^{+0.64}$ \\
2024-05-30 & 60460.42 & 00016259029 & PC & 0.77 & $1.69_{-0.18}^{+0.18}$ & $2.38_{-0.35}^{+0.41}$ \\
2024-06-06 & 60467.93 & 00016259030 & PC & 1.12 & $1.70_{-0.14}^{+0.14}$ & $2.77_{-0.31}^{+0.35}$ \\
2024-06-13 & 60474.82 & 00016259031 & PC & 0.80 & $1.75_{-0.21}^{+0.22}$ & $1.67_{-0.29}^{+0.34}$ \\
2024-07-06 & 60497.50 & 00016259033 & PC & 2.32 & $1.55$ & $0.06_{-0.03}^{+0.04}$ \\
2024-08-04 & 60526.30 & 00016259034 & PC & 0.89 & -- & $< 0.02$ \\
2024-08-11 & 60533.25 & 00016259035 & PC & 0.90 & -- & $< 0.005$ \\
2024-08-18 & 60540.30 & 00016259036 & PC & 0.66 & -- & $< 0.01$ \\
2024-08-25 & 60547.32 & 00016259037 & PC & 0.92 & -- & $< 0.02$ \\
    \end{tabular}
    \tablefoot{The spectra were fitted with an absorbed power-law model ($\texttt{tbabs$*$powerlaw}$) with a fixed hydrogen column density $N_\mathrm{H} = 2.64 \times 10^{21}~\mathrm{cm}^{-2}$. The errors are assumed to be purely statistical and are given at the $90\%$ confidence level.
    \tablefoottext{a}{XRT mode, either Photon Counting (PC) or Windowed Timing (WT).}
    \tablefoottext{b}{Unabsorbed X-ray flux in the 1--10 keV band obtained from spectral fitting, in units of $10^{-11}~\mathrm{erg~cm^{-2}~s}$.}
    }
    \label{tab:obs_swift}
\end{table*}

\clearpage
\LTcapwidth=\textwidth
\begin{longtable}{ccccc}
\caption{Paired hard-state luminosities selected for the radio/X-ray correlation. To compute $L_\mathrm{R}$ (5--GHz) and $L_\mathrm{X}$ (1--10 keV), we assumed $D = 3.9~\mathrm{kpc}$. The top section corresponds to our data, while the other sections report the luminosities computed using data from \citet{CadolleBel2007}, \citet{Soleri2010}, \citet{Tomsick2015}, \citet{Rushton2016} and \citet{Plotkin2017}. $^{(*)}$~inferred spectral index (see Sect. \ref{subsection:correlation}).}\\
\hline
MJD & $L_\mathrm{X}$ & $L_\mathrm{R}$ & $\alpha$ & $\nu$\\
    & [$\mathrm{erg~s^{-1}}$] & [$\mathrm{erg~s^{-1}}$] & & [GHz]\\
\hline
\endfirsthead
\caption{continued}\\
\hline
MJD & $L_\mathrm{X}$ & $L_\mathrm{R}$ & $\alpha$ & $\nu$\\
    & [$\mathrm{erg~s^{-1}}$] & [$\mathrm{erg~s^{-1}}$] & & [GHz]\\
\hline
\endhead
\endfoot
\endlastfoot
60223.61 & $(8.11_{-0.89}^{+1.18}) \times 10^{34}$ & $(2.07 \pm 0.32) \times 10^{28}$ & $0.35 \pm 0.04$ & 1.28 \\
60231.54 & $(4.23_{-0.09}^{+0.12}) \times 10^{35}$ & $(2.70 \pm 0.28) \times 10^{28}$ & $0.33 \pm 0.05$ & 1.28 \\
60239.49 & $(4.18_{-0.67}^{+0.77}) \times 10^{35}$ & $(3.04 \pm 0.34) \times 10^{28}$ & $0.31 \pm 0.06$ & 1.28 \\
60245.53 & $(3.71_{-0.19}^{+0.20}) \times 10^{35}$ & $(2.92 \pm 0.26) \times 10^{28}$ & $0.35 \pm 0.04$ & 1.28 \\
60254.44 & $(3.12_{-0.47}^{+0.55}) \times 10^{35}$ & $(2.99 \pm 0.26) \times 10^{28}$ & $0.34 \pm 0.06$ & 1.28 \\
60260.49 & $(2.77_{-0.12}^{+0.11}) \times 10^{35}$ & $(2.56 \pm 0.21) \times 10^{28}$ & $0.37 \pm 0.07$ & 1.28 \\
60407.98 & $(2.10_{-0.42}^{+0.52}) \times 10^{34}$ & $(1.17 \pm 0.28) \times 10^{28}$ & $0.28 \pm 0.14$ & 1.28 \\
60413.03 & $(4.11_{-0.53}^{+0.61}) \times 10^{34}$ & $(1.37 \pm 0.31) \times 10^{28}$ & $0.32 \pm 0.14$ & 1.28 \\
60421.99 & $(6.93_{-0.74}^{+0.82}) \times 10^{34}$ & $(1.40 \pm 0.37) \times 10^{28}$ & $0.36 \pm 0.15$ & 1.28 \\
60429.04 & $(8.96_{-0.93}^{+1.02}) \times 10^{34}$ & $(1.60 \pm 0.32) \times 10^{28}$ & $0.29 \pm 0.13$ & 1.28 \\
60435.94 & $(9.22_{-1.14}^{+1.29}) \times 10^{34}$ & $(1.60 \pm 0.37) \times 10^{28}$ & $0.33 \pm 0.14$ & 1.28 \\
60441.96 & $(8.57_{-0.83}^{+0.92}) \times 10^{34}$ & $(1.71 \pm 0.41) \times 10^{28}$ & $0.30 \pm 0.14$ & 1.28 \\
60448.98 & $(6.96_{-0.92}^{+1.06}) \times 10^{34}$ & $(1.57 \pm 0.30) \times 10^{28}$ & $0.42 \pm 0.12$ & 1.28 \\
60455.98 & $(5.36_{-0.88}^{+1.04}) \times 10^{34}$ & $(1.50 \pm 0.39) \times 10^{28}$ & $0.49 \pm 0.13$ & 1.28 \\
60463.95 & $(4.66_{-0.61}^{+0.70}) \times 10^{34}$ & $(1.34 \pm 0.38) \times 10^{28}$ & $0.43 \pm 0.15$ & 1.28 \\
60469.93 & $(4.62_{-0.59}^{+0.68}) \times 10^{34}$ & $(1.09 \pm 0.32) \times 10^{28}$ & $0.46 \pm 0.15$ & 1.28 \\
60476.86 & $(1.81_{-0.32}^{+0.39}) \times 10^{34}$ & $(1.06 \pm 0.25) \times 10^{28}$ & $0.49 \pm 0.15$ & 1.28 \\
60490.76 & $(2.75_{-0.95}^{+1.27}) \times 10^{33}$ & $(5.20 \pm 1.73) \times 10^{27}$ & $0.55 \pm 0.15$ & 1.28 \\
60499.73 & $(8.17_{-4.30}^{+6.13}) \times 10^{32}$ & $(3.87 \pm 1.22) \times 10^{27}$ & $0.59 \pm 0.15$ & 1.28 \\
\\
Cadolle Bel+2007\\
\hline
53593.28 & $(3.56_{-0.05}^{+0.01}) \times 10^{36}$ & $(5.92 \pm 0.64) \times 10^{28}$ & $0.03 \pm 0.03$ & 4.90 \\
\\
Soleri+2010\\
\hline
53559 & $(8.08_{-1.15}^{+1.66}) \times 10^{36}$ & $(1.59 \pm 0.17) \times 10^{29}$ & $-0.43 \pm 0.03$ & 1.40 \\
53559 & $(8.08_{-1.15}^{+1.66}) \times 10^{36}$ & $(2.49 \pm 0.04) \times 10^{29}$ & $-0.43 \pm 0.03$ & 4.80 \\
53559 & $(8.08_{-1.15}^{+1.66}) \times 10^{36}$ & $(2.23 \pm 0.05) \times 10^{29}$ & $-0.43 \pm 0.03$ & 8.40 \\
53561 & $(7.70_{-0.05}^{+0.13}) \times 10^{36}$ & $(1.39 \pm 0.22) \times 10^{29}$ & $-0.58 \pm 0.09$ & 1.40 \\
53561 & $(7.70_{-0.05}^{+0.13}) \times 10^{36}$ & $(1.99 \pm 0.12) \times 10^{29}$ & $-0.58 \pm 0.09$ & 4.80 \\
53561 & $(7.70_{-0.05}^{+0.13}) \times 10^{36}$ & $(1.40 \pm 0.15) \times 10^{29}$ & $-0.58 \pm 0.09$ & 8.40 \\
53570 & $(7.65_{-0.05}^{+0.01}) \times 10^{36}$ & $(2.61 \pm 0.38) \times 10^{29}$ & $-0.07 \pm 0.07$ & 1.40 \\
53570 & $(7.65_{-0.05}^{+0.01}) \times 10^{36}$ & $(2.28 \pm 0.05) \times 10^{29}$ & $-0.07 \pm 0.07$ & 8.40 \\
53570 & $(7.65_{-0.05}^{+0.01}) \times 10^{36}$ & $(2.16 \pm 0.12) \times 10^{29}$ & $-0.07 \pm 0.07$ & 4.80 \\
53577 & $(6.08_{-0.04}^{+0.01}) \times 10^{36}$ & $(1.67 \pm 0.50) \times 10^{29}$ & $0.26 \pm 0.14$ & 1.40 \\
53577 & $(6.08_{-0.04}^{+0.01}) \times 10^{36}$ & $(1.44 \pm 0.07) \times 10^{29}$ & $0.26 \pm 0.14$ & 8.40 \\
53577 & $(6.08_{-0.04}^{+0.01}) \times 10^{36}$ & $(1.38 \pm 0.12) \times 10^{29}$ & $0.26 \pm 0.14$ & 4.80 \\
53585 & $(4.52_{-0.04}^{+0.00}) \times 10^{36}$ & $(7.48 \pm 1.07) \times 10^{28}$ & $-0.53 \pm 0.20$ & 4.80 \\
53585 & $(4.52_{-0.04}^{+0.00}) \times 10^{36}$ & $(5.64 \pm 1.20) \times 10^{28}$ & $-0.53 \pm 0.20$ & 8.40 \\
53585 & $(4.52_{-0.04}^{+0.00}) \times 10^{36}$ & $(5.44 \pm 1.98) \times 10^{28}$ & $-0.53 \pm 0.20$ & 1.40 \\
53589 & $(3.84_{-0.03}^{+0.01}) \times 10^{36}$ & $(5.17 \pm 0.64) \times 10^{28}$ & $0.01^*$ & 8.40 \\
53593 & $(3.56_{-0.05}^{+0.01}) \times 10^{36}$ & $(6.66 \pm 0.10) \times 10^{28}$ & $-0.14 \pm 0.09$ & 8.40 \\
53593 & $(3.56_{-0.05}^{+0.01}) \times 10^{36}$ & $(6.69 \pm 0.45) \times 10^{28}$ & $-0.14 \pm 0.09$ & 4.80 \\
53593 & $(3.56_{-0.05}^{+0.01}) \times 10^{36}$ & $(6.53 \pm 1.59) \times 10^{28}$ & $-0.14 \pm 0.09$ & 1.40 \\
53624 & $(1.73_{-0.03}^{+0.01}) \times 10^{36}$ & $(3.07 \pm 0.58) \times 10^{28}$ & $0.18^*$ & 8.40 \\
53624 & $(1.73_{-0.03}^{+0.01}) \times 10^{36}$ & $(6.69 \pm 1.10) \times 10^{28}$ & $0.18^*$ & 4.80 \\
53665 & $(1.19_{-0.00}^{+0.01}) \times 10^{36}$ & $(3.03 \pm 0.55) \times 10^{28}$ & $0.22^*$ & 4.80 \\
53693 & $(9.90_{-0.05}^{+0.04}) \times 10^{35}$ & $(4.41 \pm 0.83) \times 10^{28}$ & $0.24^*$ & 4.80 \\
54282 & $(1.58_{-0.00}^{+0.02}) \times 10^{36}$ & $(2.56 \pm 0.46) \times 10^{28}$ & $0.19^*$ & 4.90 \\
54289 & $(1.54_{-0.22}^{+0.20}) \times 10^{36}$ & $(3.56 \pm 0.46) \times 10^{28}$ & $0.19^*$ & 4.90 \\
54296 & $(1.63_{-0.02}^{+0.02}) \times 10^{36}$ & $(3.83 \pm 0.46) \times 10^{28}$ & $0.18^*$ & 4.90 \\
54991 & $(1.56_{-0.01}^{+0.00}) \times 10^{36}$ & $(4.12 \pm 0.16) \times 10^{28}$ & $0.19^*$ & 8.40 \\
\\
Tomsick+2015\\
\hline
56752 & $(4.60_{-0.07}^{+0.07}) \times 10^{35}$ & $(2.00 \pm 0.09) \times 10^{28}$ & $0.29 \pm 0.05$ & 5.25 \\
\\
Rushton+2016\\
\hline
56450.62 & $(6.35_{-0.16}^{+0.17}) \times 10^{35}$ & $(1.65 \pm 0.55) \times 10^{28}$ & $0.28^*$ & 15.70 \\
56457.36 & $(7.41_{-0.10}^{+0.08}) \times 10^{35}$ & $(1.69 \pm 0.77) \times 10^{28}$ & $0.27^*$ & 15.70 \\
56461.56 & $(6.64_{-0.10}^{+0.09}) \times 10^{35}$ & $(1.47 \pm 0.81) \times 10^{28}$ & $0.28^*$ & 15.70 \\
56473.44 & $(6.09_{-0.09}^{+0.08}) \times 10^{35}$ & $(2.20 \pm 1.07) \times 10^{28}$ & $0.29^*$ & 15.70 \\
56703.79 & $(5.04_{-0.35}^{+0.29}) \times 10^{35}$ & $(1.91 \pm 0.91) \times 10^{28}$ & $0.30^*$ & 15.70 \\
56709.80 & $(4.86_{-0.20}^{+0.10}) \times 10^{35}$ & $(1.23 \pm 0.67) \times 10^{28}$ & $0.30^*$ & 15.70 \\
56724.79 & $(5.28_{-0.57}^{+0.30}) \times 10^{35}$ & $(1.74 \pm 0.68) \times 10^{28}$ & $0.30^*$ & 15.70 \\
56736.86 & $(4.82_{-0.25}^{+0.23}) \times 10^{35}$ & $(1.72 \pm 0.58) \times 10^{28}$ & $0.30^*$ & 15.70 \\
56765.41 & $(8.68_{-0.33}^{+0.03}) \times 10^{35}$ & $(2.07 \pm 0.60) \times 10^{28}$ & $0.26^*$ & 15.70 \\
56772.95 & $(1.01_{-0.02}^{+0.01}) \times 10^{36}$ & $(2.07 \pm 0.79) \times 10^{28}$ & $0.24^*$ & 15.70 \\
56779.61 & $(9.22_{-0.17}^{+0.14}) \times 10^{35}$ & $(1.82 \pm 0.62) \times 10^{28}$ & $0.25^*$ & 15.70 \\
56786.87 & $(1.17_{-0.02}^{+0.02}) \times 10^{36}$ & $(2.94 \pm 0.63) \times 10^{28}$ & $0.23^*$ & 15.70 \\
56793.13 & $(1.02_{-0.03}^{+0.02}) \times 10^{36}$ & $(3.15 \pm 0.72) \times 10^{28}$ & $0.24^*$ & 15.70 \\
56797.48 & $(9.73_{-0.18}^{+0.10}) \times 10^{35}$ & $(2.54 \pm 0.74) \times 10^{28}$ & $0.24^*$ & 15.70 \\
56799.73 & $(9.48_{-0.17}^{+0.21}) \times 10^{35}$ & $(2.22 \pm 0.75) \times 10^{28}$ & $0.25^*$ & 15.70 \\
56803.73 & $(9.01_{-0.15}^{+0.16}) \times 10^{35}$ & $(2.63 \pm 0.86) \times 10^{28}$ & $0.25^*$ & 15.70 \\
56828.54 & $(1.20_{-0.02}^{+0.02}) \times 10^{36}$ & $(3.22 \pm 0.97) \times 10^{28}$ & $0.22^*$ & 15.70 \\
56842.52 & $(9.68_{-0.22}^{+0.22}) \times 10^{35}$ & $(2.01 \pm 1.17) \times 10^{28}$ & $0.25^*$ & 15.70 \\
56856.66 & $(7.85_{-0.16}^{+0.16}) \times 10^{35}$ & $(1.90 \pm 1.02) \times 10^{28}$ & $0.26^*$ & 15.70 \\
56870.78 & $(6.84_{-0.15}^{+0.15}) \times 10^{35}$ & $(1.55 \pm 0.83) \times 10^{28}$ & $0.28^*$ & 15.70 \\
56884.37 & $(8.78_{-0.11}^{+0.15}) \times 10^{35}$ & $(1.85 \pm 0.93) \times 10^{28}$ & $0.25^*$ & 15.70 \\
56885.31 & $(8.64_{-0.10}^{+0.09}) \times 10^{35}$ & $(1.89 \pm 0.90) \times 10^{28}$ & $0.26^*$ & 15.70 \\
56752.02 & $(6.78_{-0.07}^{+0.07}) \times 10^{35}$ & $(1.44 \pm 0.62) \times 10^{28}$ & $0.28^*$ & 15.70 \\
\\
Plotkin+2017\\
\hline
57799.24 & $(9.64_{-1.18}^{+1.79}) \times 10^{34}$ & $(1.73 \pm 0.33) \times 10^{28}$ & $0.38^*$ & 15.50 \\
57803.25 & $(1.09_{-0.14}^{+0.28}) \times 10^{35}$ & $(2.07 \pm 0.29) \times 10^{28}$ & $0.37^*$ & 15.50 \\
57837.12 & $(6.15_{-0.52}^{+1.53}) \times 10^{34}$ & $(1.01 \pm 0.29) \times 10^{28}$ & $0.39^*$ & 15.50 \\
57844.10 & $(6.84_{-1.89}^{+6.22}) \times 10^{34}$ & $(1.18 \pm 0.26) \times 10^{28}$ & $0.39^*$ & 15.50 \\
57851.11 & $(3.15_{-0.45}^{+1.08}) \times 10^{34}$ & $(1.16 \pm 0.32) \times 10^{28}$ & $0.41^*$ & 15.50 \\
57862.31 & $(3.13_{-0.85}^{+1.46}) \times 10^{33}$ & $(3.03 \pm 0.47) \times 10^{27}$ & $0.44^*$ & 9.80 \\
57872.60 & $(3.87_{-1.21}^{+2.43}) \times 10^{32}$ & $(1.26 \pm 0.33) \times 10^{27}$ & $0.47^*$ & 9.80 \\
\label{tab:correlation}
\end{longtable}
\twocolumn

\section{Radio/X-ray correlation}

\begin{figure}[!ht]
    \centering
    \includegraphics[width=\linewidth]{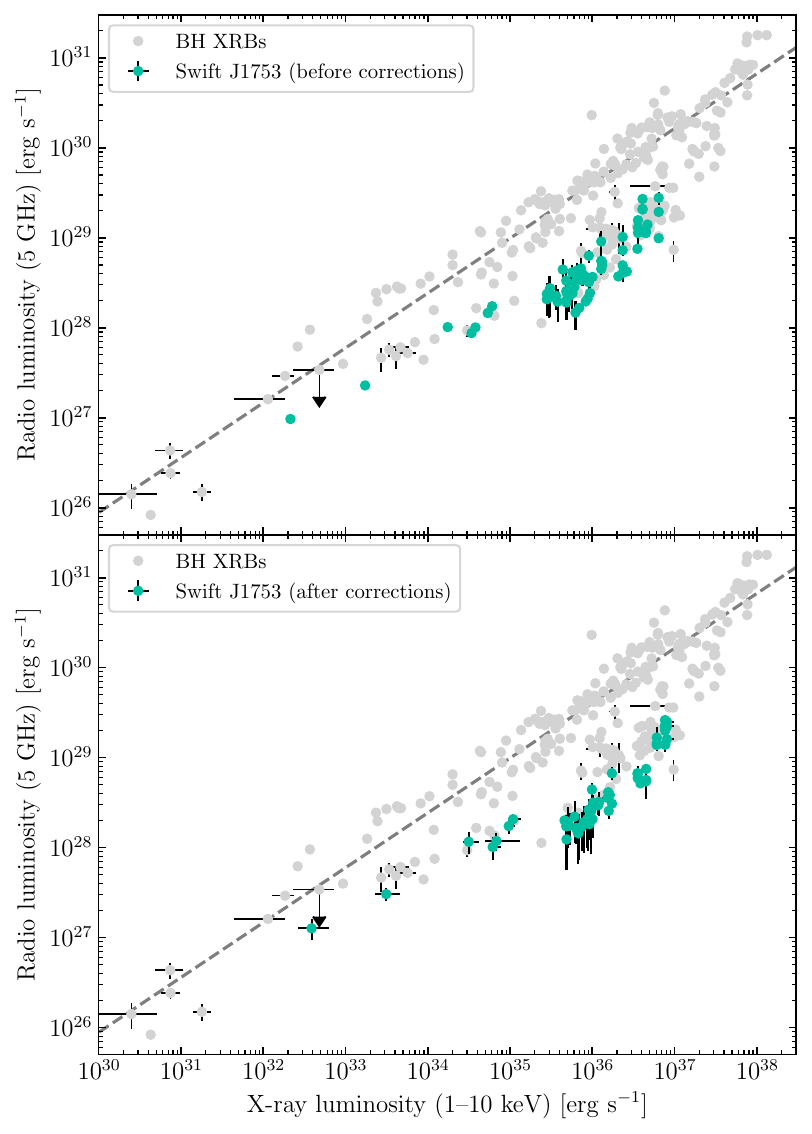}
    \caption{The radio/X-ray luminosity plane of Swift~J1753, without data from this work, rescaled at 3.9 kpc. \textit{Top panel}: Data from \citet{Soleri2010}, \citet{Rushton2016} and \citet{Plotkin2017}, compiled by \citet{Bahramian2022}. \textit{Bottom panel}: Same data, after performing the corrections detailed in Sect. \ref{subsection:correlation}.}
    \label{fig:radio_xray_corrected}
\end{figure}

Fig. \ref{fig:radio_xray_corrected} illustrates the effect of the procedure described in Sect. \ref{subsection:correlation}, aimed at reducing the scatter in the radio/X-ray plane of Swift~J1753. Here, we only show archival data, before (top panel) and after (bottom panel) correction of the luminosities.

\begin{figure}[!ht]
    \centering
    \includegraphics[width=\linewidth]{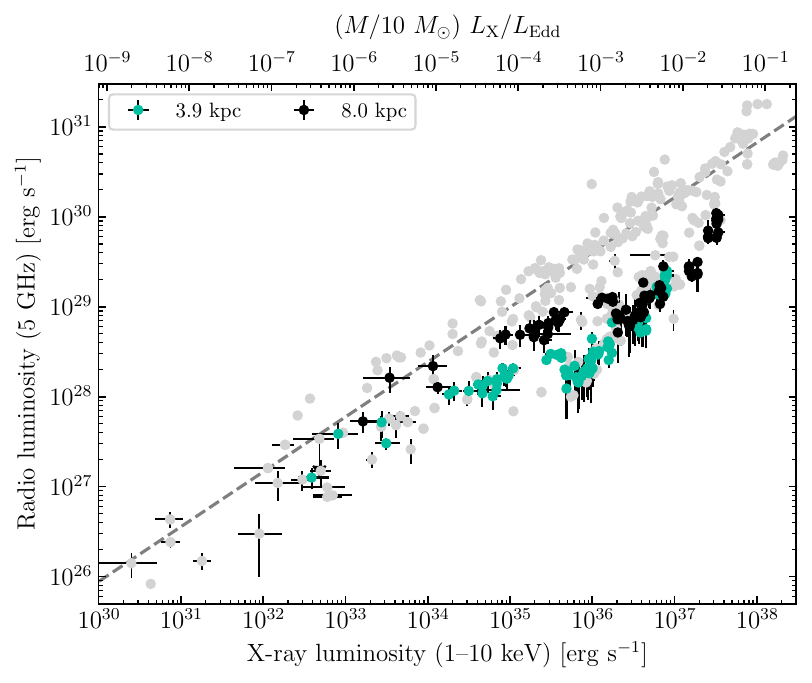}
    \caption{The radio/X-ray luminosity plane of Swift~J1753, with the same data as in Fig. \ref{fig:correlation_outl}, for two distances.}
    \label{fig:radio_xray_distance}
\end{figure}

Fig. \ref{fig:radio_xray_distance} shows the same data as in Fig. \ref{fig:correlation_outl} (archival and new data), scaled at 3.9 kpc \citep{Yanes-Rizo2025} and 8.0 kpc \citep[more consistent with previous parallax measurements; ][]{Gandhi2019, Atri2019}.

\end{appendix} 

\end{document}